\documentclass[prd,twocolumn,showpacs,amsmath,amssymb,flushbottom]{revtex4-1}
\usepackage{graphicx}
\usepackage{epstopdf}
\usepackage[colorlinks,citecolor=blue,linkcolor=blue]{hyperref}
\usepackage{enumerate}
\usepackage{dcolumn}

\usepackage{relsize}
\def\babar{\mbox{\slshape B\kern-0.1em{\smaller A}\kern-0.1em
    B\kern-0.1em{\smaller A\kern-0.2em R}}}

\numberwithin{equation}{section}

\begin{document}

\title{\boldmath
Study of Dynamics of $D^0 \to K^- e^+ \nu_{e}$ and $D^0\to\pi^- e^+ \nu_{e}$ Decays
}
\author{
  \begin{small}
    \begin{center}
      M.~Ablikim$^{1}$, M.~N.~Achasov$^{9,f}$, X.~C.~Ai$^{1}$,
      O.~Albayrak$^{5}$, M.~Albrecht$^{4}$, D.~J.~Ambrose$^{44}$,
      A.~Amoroso$^{49A,49C}$, F.~F.~An$^{1}$, Q.~An$^{46,a}$,
      J.~Z.~Bai$^{1}$
% ...
      \\
      \vspace{0.2cm}
      (BESIII Collaboration)\\
      \vspace{0.2cm} {\it
        $^{1}$ Institute of High Energy Physics, Beijing 100049, People's Republic of China\\
% ...
$^{4}$ Bochum Ruhr-University, D-44780 Bochum, Germany\\
        $^{5}$ Carnegie Mellon University, Pittsburgh, Pennsylvania 15213, USA\\
% ...
$^{9}$ G.I. Budker Institute of Nuclear Physics SB RAS (BINP), Novosibirsk 630090, Russia\\
% ...
$^{44}$ University of Rochester, Rochester, New York 14627, USA\\
% ...
$^{46}$ University of Science and Technology of China, Hefei 230026, People's Republic of China\\
% ...
$^{49}$ (A)University of Turin, I-10125, Turin, Italy; (B)University of Eastern Piedmont, I-15121, Alessandria, Italy; (C)INFN, I-10125, Turin, Italy\\
% ...
$^{a}$ Also at State Key Laboratory of Particle Detection and Electronics, Beijing 100049, Hefei 230026, People's Republic of China\\
% ...
$^{f}$ Also at the Novosibirsk State University, Novosibirsk, 630090, Russia\\
% ...
      }
    \end{center}
    \vspace{0.4cm}
  \end{small}
}
\affiliation{}


\begin{abstract}
In an analysis of a 2.92~fb$^{-1}$ data sample taken at 3.773~GeV with
the BESIII detector operated at the BEPCII collider,
we measure the absolute decay branching fractions to be
$\mathcal B(D^0 \to K^-e^+\nu_e)=(3.505\pm 0.014 \pm 0.033)\%$ and
$\mathcal B(D^0 \to \pi^-e^+\nu_e)=(0.295\pm 0.004\pm 0.003)\%$.
From a study of the differential decay rates
we obtain the products of hadronic form factor and
the magnitude of the CKM matrix element
$f_{+}^K(0)|V_{cs}|=0.7172\pm0.0025\pm 0.0035$ and
$f_{+}^{\pi}(0)|V_{cd}|=0.1435\pm0.0018\pm 0.0009$.
Combining these products with the values of $|V_{cs(d)}|$
from the SM constraint fit,
we extract the hadronic form factors
$f^K_+(0)   = 0.7368\pm0.0026\pm 0.0036$ and
$f^\pi_+(0) = 0.6372\pm0.0080\pm 0.0044$,
and their ratio
$f_+^{\pi}(0)/f_+^{K}(0)=0.8649\pm 0.0112\pm 0.0073$.
These form factors and their ratio
are used to test unquenched Lattice QCD calculations of the form factors
and a light cone sum rule (LCSR) calculation of their ratio.
The measured value of $f_+^{K(\pi)}(0) |V_{cs(d)}|$ and
the lattice QCD value for $f^{K(\pi)}_+(0)$
are used to extract values of the CKM matrix elements of
$|V_{cs}|=0.9601 \pm 0.0033 \pm 0.0047 \pm 0.0239$ and
$|V_{cd}|=0.2155 \pm 0.0027 \pm 0.0014 \pm 0.0094$,
where the third errors are due to the uncertainties in lattice QCD calculations of the form factors.
Using the LCSR value for $f_+^\pi(0)/f_+^K(0)$, we determine the ratio
$|V_{cd}|/|V_{cs}|=0.238\pm 0.004\pm 0.002\pm 0.011$,
where the third error is from the uncertainty in the LCSR normalization.
In addition, we measure form factor parameters for
three different theoretical models that describe the weak hadronic charged currents
for these two semileptonic decays.
All of these measurements are the most precise to date.
\end{abstract}

\pacs{13.20.Fc, 12.15.Hh}

\maketitle

\section{Introduction}
\label{sec:intr}

In the Standard Model (SM) of particle physics, the mixing between the quark flavors
in the weak interaction is parameterized by the unitary $3\times3$
Cabibbo-Kobayashi-Maskawa (CKM) matrix $\hat V_{\rm CKM}$~\cite{Cabibbo,KM}.
The CKM matrix elements are fundamental parameters of the SM, which have to be measured in experiments.
Beyond the SM, some New Physics (NP) effects would also be involved in the weak interactions of the quark flavors,
and modify the coupling strength of the quark flavor transitions.
Due to these two reasons, precise measurements of the CKM matrix elements are very important for many tests of the SM
and searches for NP beyond the SM.
Each CKM matrix element can be extracted from measurements of different processes
supplemented by theoretical calculations of corresponding hadronic matrix elements.
Since the effects of the strong and weak interactions can be well separated in
semileptonic $D^0 \to K^- e^+\nu_e$ and $D^0 \to \pi^- e^+\nu_e$ decays,
these processes are well suited for the determination of
the magnitudes of the CKM matrix elements $V_{cs}$ and $V_{cd}$,
and also for studies of the weak decay mechanisms of charmed mesons.
If any significant inconsistency between the precise direct measurements of $|V_{cd}|$
or $|V_{cs}|$ and those obtained from the SM global fit is observed, it may indicate that some NP effects
are involved in the first two quark generations~\cite{GRong_CPC34_788}.

In the limit of zero positron mass, the differential rate for
$D^0 \to K^- (\pi^-) e^+\nu_e$ decay is given by
\begin{equation}
\frac {d\Gamma }{dq^2} = \frac {G_F^2}{24\pi ^3}|V_{cs(d)}|^2
|\vec p_{K^-(\pi^-)}|^3
|f_+^{K(\pi)}(q^2)|^2,
\label{eq_dGamma_dq2}
\end{equation}
where $G_F$ is the Fermi coupling constant,
$\vec p_{K^-(\pi^-)}$ is the three-momentum of the
$K^-$($\pi^-$) meson in the rest frame of the $D^0$ meson,
and $f^{K(\pi)}_+(q^2)$ represents the
hadronic form factors of the hadronic weak current that depend on
the square of the four-momentum transfer $q = p_{D^0} - p_{K^-(\pi^-)}$.
These form factors describe strong interaction effects that
can be calculated in lattice quantum chromodynamics (LQCD).

In recent years, LQCD has provided calculations of these form factors with steadily increasing precision.
With these improvements in precision,
experimental validation of the computed results are more and more important.
At present, the main uncertainty of the apex of the $B_d$ unitarity triangle (UT) of $B$ meson decays is dominated by the theoretical errors in the LQCD determinations of the
$B$ meson decay constants $f_{B_{(s)}}$ and decay form factors
$f_+^{B \rightarrow \pi}(0)$~\cite{GRong_CPC34_788}.
Precision measurements of the charmed-sector
form factors $f^{K(\pi)}_+(q^2)$ can be used to
establish the level of reliability of
LQCD calculations of $f_+^{B \rightarrow \pi}(0)$.
If the LQCD calculations of $f^{K(\pi)}_+(q^2)$ agree well with
measured $f^{K(\pi)}_+(q^2)$ values,
the LQCD calculations of the form factors for $B$ meson semileptonic decays
can be more confidently used to improve measurements of $B$ meson semileptonic decay rates.
The improved measurements of $B$ meson semileptonic decay rates would,
in turn, improve the
determination of the $B_d$ unitarity triangle, with which
one can more precisely test the SM
and search for NP.

In this paper, we present direct measurements of the absolute branching fractions
for $D^0 \to K^- e^+\nu_e$ and $D^0 \to \pi^- e^+\nu_e$ decays
using a 2.92~fb$^{-1}$ data sample
taken at 3.773 GeV with
the BESIII detector~\cite{bes-iii} operated at the upgraded Beijing Electron Positron
Collider (BEPCII)~\cite{BEPCII} during the time period
from 2010 to 2011.
(Throughout this paper, the inclusion of charge conjugate channels is implied.)
By analyzing partial decay rates for
$D^0 \to K^- e^+\nu _e$ and $D^0 \to \pi^- e^+\nu _e$,
we obtain the $q^2$ dependence of the form factors $f^{K(\pi)}_+(q^2)$.
Furthermore we
extract the form factors $f_+^K(0)$ and $f_+^\pi(0)$
using values of $|V_{cs}|$ and $|V_{cd}|$
determined by the CKMfitter group~\cite{pdg2014}.
Conversely, taking LQCD values for $f^K_+(0)$ and $f^\pi_+(0)$ as inputs,
we determine the values of the CKM matrix elements $|V_{cs}|$ and  $|V_{cd}|$.

We review the approaches for describing the dynamics
of $D^0 \rightarrow K^-e^+\nu_e$ and $D^0 \rightarrow \pi^-e^+\nu_e$ decays in Section~\ref{sec:th}.
We then describe the BESIII detector, the data sample and the simulated Monte Carlo events
used in this analysis in Section~\ref{sec:bes3}. In Section~\ref{sec:rec}, we introduce the analysis technique used to identify
the semileptonic decay events. The measurements of the absolute branching fractions for these two decays
and study of systematic uncertainties in these branching fraction measurements are described in Section~\ref{sec:bf}.
In Section~\ref{sec:dr}, we describe the analysis techniques for measuring the differential decay rates
for these two semileptonic decays, and present our measurements of the
hadronic form factors. The determinations of the CKM matrix elements
$|V_{cs}|$ and $|V_{cd}|$
are discussed in Section~\ref{sec:vcq}. We give a summary of our measurements
in Section~\ref{sec:sum}.

\section{\boldmath Form factor and approaches for $D^0$ Semileptonic Decays}
\label{sec:th}

\subsection{Hadronic form factor}

In general, the form factor $f_+^{K(\pi)}(q^2)$ can be expressed in terms
of a dispersion relation~\cite{form_factor_dispersion_relation}
\begin{equation}
     f_+(q^2)=  \frac   {f_+(0)/{(1-\alpha)} } {1- \frac{q^2}{M^2_{\rm pole}} }
              + \frac{1}{\pi}\int_{t_+}^{\infty}dt \frac{{\rm Im}f_+(t)}{t-q^2-i\epsilon},
\label{form_factor_dispersion_relation}
\end{equation}
where $M_{\rm pole}$ is the mass of the lowest-lying relevant vector meson,
for $D^0\rightarrow K^- e^+\nu_e$ it is the $D^{*+}_s$,
while for $D^0\rightarrow \pi^- e^+\nu_e$ it is the $D^{*+}$,
$f_+(0)$ is the form factor evaluated at the four-momentum transfer $q=0$,
$\alpha$ is the relative size of the contribution to $f_+(0)$ from the vector pole at $q^2=0$,
$t_+=(m_{D^0}+m_{K^-(\pi^-)})^2$ corresponds to the threshold for
$D^0K^-(\pi^-)$ production,
$m_{D^0}$ and $m_{K^-(\pi^-)}$ are the masses of the $D^0$ and charged kaon (pion) meson, respectively.
From Eq.~(\ref{form_factor_dispersion_relation}) we find that, except
for the pole position of
the lowest-lying meson being located below threshold, $f_+(q^2)$
is analytic outside of a cut in the complex $q^2$-plane extending along the real axis
from ${t_+}$ to ${\infty}$, corresponding to the production region for the states
with the appropriate quantum numbers.

\subsection{Parameterizations of form factor}

The form of the dispersion relation given in Eq.~(\ref{form_factor_dispersion_relation})
is often parameterized by keeping the lowest-lying meson pole explicitly
and approximating the remaining dispersion integral in
Eq.~(\ref{form_factor_dispersion_relation})
by a number of effective poles~\cite{form_factor_dispersion_relation, PLB_478_417}
\begin{equation}
     f_+(q^2)=  \frac   {f_+(0)/{(1-\alpha)} } {1- \frac{q^2}{M^2_{\rm pole}} }
              + \frac{1}{\pi} \sum_{k=1}^{N} \frac{\rho_k} {1-\frac{1}{\gamma_k} \frac{q^2}{M^2_{\rm pole}}},
\label{form_factor_dispersion_relation_2}
\end{equation}
where ${\rho_k}$ and $\gamma_k$ are expansion parameters that are not predicted.
The form factor can be approximated by introducing arbitrarily many effective poles.
Equation (\ref{form_factor_dispersion_relation_2}) is the starting point for many proposed
form factor parameterizations.

\subsubsection{Single pole form}
In the constituent quark model, lattice gauge calculations, and QCD sum rules,
such as the K\"orner-Schuler (KS)~\cite{ks} and Bauer-Stech-Wirbel (BSW)~\cite{wsb} models,
a commonly used form factor has a single pole of the form
\begin{equation}
     f_+(q^2)=\frac{f_+(0)}{1-\frac{q^2}{M^2_{\rm pole}}},
\label{pole}
\end{equation}
which is simply the first term in Eq.~(\ref{form_factor_dispersion_relation_2})
(taking $\alpha=0$).
The pole mass $M_{\rm pole}$ is often treated as a free parameter to improve fit quality.

\subsubsection{Modified pole model}

The modified pole model uses $N=1$ in Eq.~(\ref{form_factor_dispersion_relation_2}), and
 the form factor can be expressed as
\begin{equation}
     f_+(q^2)=\frac{f_+(0)}{(1-\frac{q^2}{M^2_{\rm pole}})(1-\alpha \frac{q^2}{M^2_{\rm pole}})},
\label{modifies-pole_model}
\end{equation}
where $\alpha$ is a free parameter.
This model is the so-called Becirevic-Kaidalov (BK)
parameterization~\cite{PLB_478_417} and has been
used in many recent lattice calculations and experimental studies for
$D^0$ meson semileptonic decays.

\subsubsection{Series expansion}

The series expansion~\cite{form_factor_dispersion_relation}
is the most general parameterization that is consistent with constraints from QCD.
It has the form
\begin{equation}
     f_{+}(t)=\frac{1}{P(t)\Phi(t,t_0)} a_0(t_0)\left(1+\sum_{k=1}^{\infty} r_k(t_0)[z(t,t_0)]^k \right),
 \label{series_expansion_model}
\end{equation}
where
\begin{align}
z(t,t_0) &= \frac{\sqrt{t_+-t}-\sqrt{t_+-t_0}}{\sqrt{t_+-t}+\sqrt{t_+-t_0}}, \\
 t_{-}   &= (m_{D^0}- m_{K^-(\pi^-)})^2, \\
 t_0     &= t_+(1-\sqrt{1-t_-/t_+}),
\end{align}
$a_0(t_0)$ and $r_k(t_0)$ are real coefficients.
The function $P(t)=z(t,m^2_{D^*_s})$ for $D\to K$  and $P(t)=1$ for $D\to \pi$.
$\Phi$ is given by
\begin{eqnarray}
\lefteqn{
 \Phi(t,t_0)  =  \sqrt{\frac{1}{24\pi\chi_V}}(\frac{t_+-t}{t_+-t_0})^{1/4}(\sqrt{t_+-t}+\sqrt{t_+})^{-5}
 }
 \nonumber \\
 && {}
 \times (\sqrt{t_+-t}+\sqrt{t_+-t_0})(\sqrt{t_+-t}+\sqrt{t_+-t_-})^{3/2}
 \nonumber \\
 && {}
 \times (t_+-t)^{3/4},
\end{eqnarray}
where $\chi_V$ can be obtained from dispersion relations using perturbative QCD and depends on
the ratio of the $s$ quark mass to the $c$ quark mass, $\xi=m_s/m_c$~\cite{Nucl_Phys_B461_493}.
At leading order, with $\xi=0$,
\begin{equation}
 \chi_V = \frac{3}{32\pi^2 m^2_c}.
\end{equation}
The choice of $P$ and $\Phi$ is such that
\begin{equation}
  a_0^2 (t_0) \left(1+\sum_{k=1}^{\infty} r_k^2(t_0)\right) \leq 1.
\end{equation}
      The $z$ series expansion is model independent and satisfies analyticity and unitarity.
In heavy quark effective theory~\cite{heavy_quark_effective_theory}
the coefficients $r_k$ in Eq.~(\ref{series_expansion_model}) for $D \rightarrow \pi e^+\nu_e$
and $B\rightarrow \pi \ell^+\nu_\ell$ decays are related.
A measurement of the $r_k$ for the decay of $D\rightarrow \pi e^+\nu_e$
therefore provides important information to constrain the class of form factors needed to fit the decays of
$B\rightarrow \pi \ell^+\nu_\ell$,
and thereby provides improvements in
the determination of the magnitude of the CKM matrix element  $V_{ub}$.
However, the validity of the form factor parameterization given in Eq.~(\ref{series_expansion_model})
still needs to be checked with experimental data.
This is one of the reasons why it is important to precisely measure the
form factors $f_+^{K(\pi)}(q^2)$ for $D^0 \rightarrow K^-(\pi^-)e^+\nu_e$ decays.

In practical applications, one often takes
$k_{\rm max}=1$ or $k_{\rm max}=2$
in Eq.~(\ref{series_expansion_model}), which gives following two forms of the form factor:
\begin{enumerate}[(a)]
  \item
  {\bf\em Series expansion with 2 parameters} of the form factor is given by
\begin{eqnarray}
     f_{+}(t)&=&\frac{1}{P(t)\Phi(t,t_0)} a_0(t_0)
     \nonumber\\
     & \times &  \left(1 + r_1(t_0)[z(t,t_0)] \right),
\end{eqnarray}
which gives
\begin{eqnarray}
     f_{+}(t)&=&\frac{1}{P(t)\Phi(t,t_0)} \frac{f_+(0)P(0)\Phi(0,t_0)}{1+r_1(t_0)z(0,t_0)} \nonumber \\
     & \times &  \left(1+ r_1(t_0)[z(t,t_0)] \right).
\end{eqnarray}

  \item
  {\bf\em Series expansion with 3 parameters} of the form factor is given by
  \begin{eqnarray}
  f_{+}(t) & = & \frac{1}{P(t)\Phi(t,t_0)} a_0(t_0)
  \nonumber\\
  &\times& (1 + r_1(t_0)[z(t,t_0)]
  \nonumber\\
  &+&  r_2(t_0)[z(t,t_0)]^2),
\end{eqnarray}
which gives
  \begin{eqnarray}
  f_{+}(t) & = &\frac{1}{P(t)\Phi(t,t_0)}
  \nonumber\\
  &\times& \frac{f_+(0)P(0)\Phi(0,t_0)}{1+r_1(t_0)z(0,t_0)+r_2(t_0)[z(0,t_0)]^2}
  \nonumber\\
  &\times& (1 + r_1(t_0)[z(t,t_0)]
  \nonumber\\
  && + r_2(t_0)[z(t,t_0)]^2).
\end{eqnarray}
\end{enumerate}

\section{Data Sample and the BESIII Experiment}
\label{sec:bes3}

At $\sqrt{s}=3.773$~GeV, the $\psi(3770)$ resonance is directly produced via $e^+e^-$ annihilation.
About $93\%$~\cite{pdg2014} of $\psi(3770)$ decays to $D\bar D$ ($D^0 \bar D^0$, $D^+D^-$) meson pairs.
In addition, the continuum processes $e^+e^-\rightarrow q\bar q$ ($q=u,d,s$ quark),
$e^+e^-\rightarrow \tau^+\tau^-$, $e^+e^-\rightarrow \gamma_{\rm ISR} J/\psi$,
$e^+e^-\rightarrow \gamma_{\rm ISR} \psi(3686)$ events are also produced,
where $\gamma_{\rm ISR}$ is the radiative photon in the initial state.
The data sample contains a mixture of all these classes of events.
In the analysis, we refer to events other than $\psi(3770)$ decays to $D\bar D$
as ``non-$D\bar D$ process'' events.

BEPCII~\cite{BEPCII} is a double-ring $e^+e^-$ collider
operating in the center-of-mass energy region between 2.0 and 4.6~GeV.
Its design luminosity at 3.78~GeV is $10^{33}\, \mathrm{cm^{-2}s^{-1}}$
with a beam current of 0.93 A.
The peak luminosity of the machine reached $0.65\, \times 10^{33}\,
\mathrm{cm^{-2}s^{-1}}$
at $\sqrt{s}=3.773$~GeV in April 2011 during the $\psi(3770)$ data taking.
BESIII~\cite{bes-iii} is a general  purpose detector operated at the BEPCII.
At the BEPCII colliding point, the $e^+$ and $e^-$ beams collide
with a crossing angle of 22 mrad.

The BESIII detector is a cylindrical magnetic detector
with a solid angle coverage of $93\%$ of $4\pi$.
It consists of several main components. Surrounding the beam pipe, there is a
43-layer main drift chamber (MDC)
that provides precise measurements
of charged particle trajectories and ionization energy losses ($dE/dx$)
that are used for particle identification.
The momentum resolution for charged particles at 1~GeV$/c$ is 0.5\%,
and the specific $dE/dx$ resolution is 6\%.
Outside of the MDC, a time-of-flight (TOF) system is
used for charged particle identification.
The TOF consists of a barrel part made of two layers with 88 pieces of
2.4~m long plastic scintillators in each layer,
and two end-caps with 96 fan-shaped detectors.
The TOF time resolution is 80 ps in the barrel, and 110 ps in the end-caps,
corresponding to a $K/\pi$ separation better than $2\sigma$ for momenta up to 1~GeV$/c$.
An electromagnetic calorimeter (EMC) surrounds the TOF
and is made of 6240 CsI(Tl) crystals arranged in a cylindrical shape (barrel) plus two end-caps.
The EMC is used to measure the energies of photons and electrons.
For 1.0~GeV photons, the energy resolution is $2.5\%$ in the barrel and $5.0\%$ in the end-caps,
and the one-dimensional position resolution is 6~mm in the barrel and 9~mm in the end-caps.
A superconducting solenoid magnet outside the EMC provides a 1~T
magnetic field in the central tracking region of the detector.
A muon identification system is placed outside of the detector,
consisting of about 1272~m$^2$
of resistive plate chambers arranged in 9 layers in the barrel and
8 layers in the end-caps incorporated
in the magnetic flux return iron of the magnetic.
The position resolution of the muon chambers is about 2~cm.
This system efficiently identifies
muons with momentum greater than 500~MeV$/c$
over $88\%$ of the total solid angle.

The BESIII detector response was studied using Monte Carlo event samples
generated with a {\sc geant4}-based~\cite{geant4} detector simulation software
package, {\sc boost}~\cite{BOOST}.
To match the data, $1.98\times 10^8$
Monte Carlo events for
$e^+e^- \rightarrow \psi(3770)\rightarrow D \bar D$ were simulated with the
Monte Carlo event generator $\mathcal{KK}$, {\sc kkmc}~\cite{kkmc},
where $56\%$ of the $\psi(3770)$ resonance
is set to decay to $D^0\bar D^0$
while the remainder decays to $D^+D^-$ meson pairs.
All of these $D^0 \bar D^0$ and $D^+D^-$ meson pairs
are set to decay into different final states
which were generated with {\sc EvtGen}~\cite{besevtgen}
with branching fractions from the Particle Data Group (PDG)~\cite{pdg2014}.
This Monte Carlo event sample corresponds to about 11 times the
luminosity of real data.
With these Monte Carlo events, we determine the event selection criteria
for the data analysis and study possible background events for the measurement
of the $D^0 \to K^-e^+\nu _e$ and $D^0 \to \pi ^-e^+\nu _e$ decays.
We refer to these Monte Carlo events
as ``cocktail vs. cocktail $D\bar D$ process'' events.

Since the ``non-$D \bar D$ process'' events are mixed with the $D\bar D$
events in the data sample,
we also generate ``non-$D \bar D$ process'' Monte Carlo events
simulated with  {\sc kkmc}~\cite{kkmc} and {\sc EvtGen}~\cite{besevtgen}
to estimate the number
of the background events in the selected $D^0 \rightarrow K^-e^+\nu_e$
and $D^0 \rightarrow \pi^-e^+\nu_e$ samples.

To estimate the efficiencies, we also generate ``Signal'' Monte Carlo events, \emph{i.e.}
$\psi(3770)\rightarrow D^0\bar D^0$ events in which the $\bar D^0$ meson
decays to all possible final states~\cite{pdg2014},
and the $D^0$ meson decays to a semileptonic
or a hadronic decay final state that is being investigated.
These Monte Carlo events were all generated and simulated with the software packages mentioned above.

\section{\boldmath Reconstruction of $D^0\bar D^0$ Decay Events}
\label{sec:rec}

In $\psi(3770)$ resonance decays into $D\bar D$ mesons in which a $\bar D$ meson is fully reconstructed,
all of the remaining tracks and photons in the event must originate from the accompanying $D$.
In these cases, the reconstructed meson
is called a single $\bar D$ tag.
Using the single $\bar D^0$ tag sample, the
decays of $D^0 \to K^-e^+\nu _e$ and $D^0 \to \pi ^-e^+\nu _e$
can be reliably identified from the recoiling tracks in the event.
We refer to the event in which the $\bar D^0$ meson is reconstructed and
a semileptonic $D^0$ decay is reconstructed from the recoiling tracks
as a doubly tagged $D^0 \bar D^0$ decay event or
a double $D^0 \bar D^0$ tag. With these doubly tagged $D^0 \bar D^0$ events,
the absolute branching fractions and the differential decay rates
for $D^0$ semileptonic decays can be well measured.

In the analysis, all 4-momentum vectors measured in the laboratory frame
are boosted to the $e^+e^-$ center-of-mass  frame.

In this section, we describe the procedure
for selecting the single $\bar D^0$ tags
and the $D^0$ semileptonic decay events.

\subsection{\boldmath Properties of doubly tagged $D^0 \bar D^0$ decays}

For a specific tag decay mode,
the number of the single $\bar D^0$ tags is given by
\begin{equation}
 N_{\rm tag} = 2  N_{D^0\bar D^0}
     {\mathcal B}_{\rm tag}  \epsilon_{\rm tag},
\end{equation}
where $N_{D^0\bar D^0}$ is the number of the $D^0\bar D^0$ meson pairs produced in the data sample,
$\mathcal B_{\rm tag}$ is the branching fraction for the tag mode,
and $\epsilon_{\rm tag}$ is the efficiency for reconstruction of this mode.
Similarly, the number of the $D^0$ semileptonic decay events observed in the system recoiling
against the single $\bar D^0$ tags is given by
\begin{eqnarray}
   N_{\rm observed}(D^0\to h^-e^+\nu_e) & = & 2  N_{D^0\bar D^0} {\mathcal B}_{\rm tag} \mathcal B(D^0\to h^-e^+\nu_e)
   \nonumber \\
   & \times & \epsilon_{{\rm tag}, D^0\to h^-e^+\nu_e},
\end{eqnarray}
where $h^-$ denotes the final state hadron (i.e. $h^-=K^-$ or $\pi^-$),
$\mathcal B(D^0\rightarrow h^-e^+\nu_e)$ is the $D^0$ meson semileptonic decay branching fraction,
and $\epsilon_{{\rm tag}, D^0\to h^-e^+\nu_e}$ is the efficiency of simultaneously reconstructing both the single
$\bar D^0$ tag and the $D^0$ meson semileptonic decay.
With these two equations, we obtain
\begin{equation}
 \mathcal B(D^0\rightarrow h^-e^+\nu_e) =
        \frac{ N_{\rm observed}(D^0 \rightarrow h^-e^+\nu_e) }
        { N_{\rm tag} \epsilon(D^0\rightarrow h^-e^+\nu_e) },
\label{bf_fotmula}
\end{equation}
where $\epsilon(D^0\rightarrow h^-e^+\nu_e)={\epsilon_{{\rm tag}, D^0\to h^-e^+\nu_e}}/{\epsilon_{\rm tag}}$.

To measure the $D^0$ semileptonic differential decay rates given in Eq.(\ref{eq_dGamma_dq2}) we need to evaluate
the partial decay rate $\Delta \Gamma_i$ observed within a small range of the squared four-momentum transfer $\Delta q^2_i$,
where $i$ stands for the $i$th $q^2$ bin.
This partial decay rate can be evaluated with the double tag $D^0 \bar D^0$ events as well.
The measurement of the partial decay rates is described in Section~\ref{sec:dr}.

\subsection{\boldmath Single $\bar D^0$ tags and efficiencies}

   The $\bar D^0$ meson is reconstructed in five
hadronic decay modes:
$K^+\pi^-$,
$K^+\pi^-\pi^0$,
$K^+\pi^-\pi^-\pi^+$,
$K^+\pi^-\pi^+\pi^-\pi^0$ and
$K^+\pi^-\pi^0\pi^0$.
Events that contain at least
two reconstructed charged tracks with good helix fits are selected.
The charged tracks used in the single tag analysis
are required to satisfy $|\cos\theta|<0.93$,
where $\theta$ is the polar angle of the charged track.
All of these charged tracks are required to
originate from the interaction region with a distance of
closest approach in the transverse plane that is less than 1.0~cm and
less than 15.0~cm along the $z$ axis.
The $dE/dx$ and TOF measurements are combined to form confidence levels
for pion ($CL_\pi$) and kaon ($CL_K$) particle identification hypotheses.
In the selection of single $\bar D^0$ tags,
pion (kaon) identification requires
$CL_{\pi}>CL_{K}$ ($CL_{K}>CL_{\pi}$) for momenta $p<0.75$~GeV/$c$
and $CL_{\pi}>0.1\%$ ($CL_{K}>0.1\%$) for $p\ge 0.75$~GeV/$c$.

A $\pi^0$ meson is reconstructed via the decay $\pi^0 \rightarrow \gamma\gamma$.
To select photons from $\pi^0$ decays, we require an
energy deposit in the barrel (end-cap) EMC
to be greater than $0.025~(0.050)$~GeV
and in-time coincidence with the beam crossing.
In addition, the angle between the
photon and the nearest charged track is required to be greater than $10^{\circ}$.
A  one-constraint (1-C) kinematic fit
is performed to constrain the invariant mass of $\gamma\gamma$ to the mass of $\pi^0$ meson,
and $\chi^2<50$ is required.

For the $\bar D^0\to K^+\pi^-$ final state, we
reduce backgrounds from cosmic rays, Bhabha and dimuon events by requiring
the difference of the time-of-flight of the two charged
tracks be less than 5~ns, and the opening angle of the two charged track directions
be less than 176 degree. In addition, we require that the sum of the ratio of
energy over the momentum
of the charged track is less than 1.4.

The single $\bar D^0$ tags are selected using beam energy constrained mass
of the $Kn\pi$ (where $n =$ 1, 2, 3, or 4) combination, which is given by
\begin{equation}
    M_{\rm BC} = \sqrt {E_{\rm beam}^2-|\vec {p}_{Kn\pi}|^2},
\end{equation}
where $E_{\rm beam}$ is the beam energy
and
$|\vec p_{Kn\pi}|$ is the magnitude of the momentum
of the daughter $Kn\pi$ system.
We also use the variable $\Delta E \equiv E_{Kn\pi}-E_{\rm beam}$,
where $E_{Kn\pi}$ is the energy
of the $Kn\pi$ combination computed with the identified charged species.
Each $Kn\pi$ combination is subjected to a requirement of energy conservation
with $|\Delta E|<(2\sim3)\sigma_{E_{Kn\pi}}$,
where
$\sigma_{E_{Kn\pi}}$ is the standard deviation of the $E_{Kn\pi}$ distribution.
For each event, there may be several different charged track (or both
charged track and neutral cluster) combinations for each of
the five single $\bar D^0$ tag modes.
If more than one combination satisfies the energy requirement,
the combination with the smallest value of $|\Delta E|$ is retained.

\begin{figure}[!hbp]
\centerline{
\includegraphics[width=0.5\textwidth]{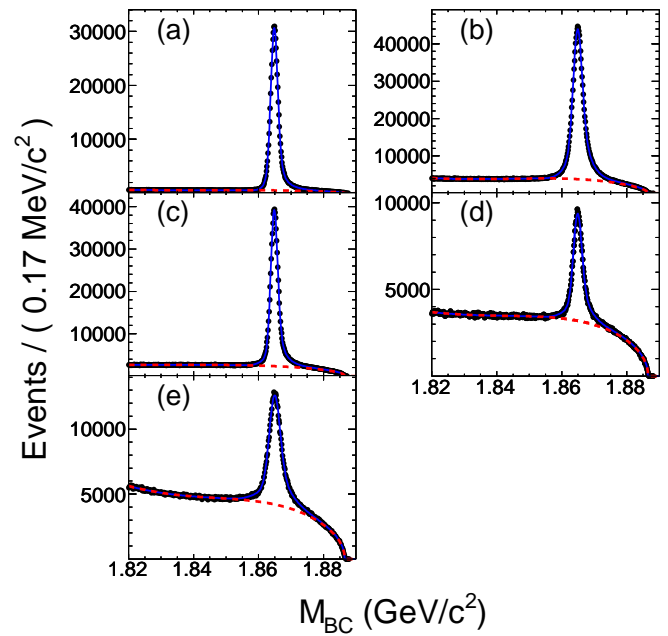}
}
\caption{Distributions of the beam energy constrained masses of the $Kn\pi$ ($n$ = 1, 2, 3 or 4)
combinations for the 5 single $\bar D^0$
tag modes: (a) $K^+\pi^-$, (b) $K^+\pi^-\pi^0$, (c) $K^+\pi^-\pi^-\pi^+$,
(d) $K^+\pi^-\pi^-\pi^+\pi^0$ and (e) $K^+\pi^-\pi^0\pi^0$.
}
\label{Fig:d_mass_bc_fit}
\end{figure}

The dots with error bars in Fig.~\ref{Fig:d_mass_bc_fit} show
the resulting distributions of $M_{\rm BC}$
for the five single $\bar D^0$ tag modes, where
the $\bar D^0$ meson signals are evident.
To determine the number of the single $\bar D^0$ tags that are reconstructed
for each mode, we fit
a signal function plus a background shape to
these distributions.
For the fit, we use signal shapes obtained from simulation
convolved with a double-Gaussian function for the signal component, added to an
ARGUS function multiplied by a third-order polynomial function~\cite{bes2_plb597_p39_2004,bes2_D_physics_papers}
to represent the combinatorial background shape.
The ARGUS function is \cite{argus_function}
\begin{equation}
f_{\rm ARGUS}(m) =  m \sqrt{ 1-\left( \frac{m}{E_0} \right)^2 }
                    \exp{ \left[ c \left (1- \left (\frac{m}{E_0} \right)^2\right) \right]},
\end{equation}
where $m$ is the beam energy constrained mass, $E_0$ is the endpoint
given by the beam energy and $c$ is a free parameter.
The  solid lines in Fig. \ref{Fig:d_mass_bc_fit} show the best fits, while the  dashed
lines show the fitted background shapes.

In addition to the combinatorial background, there are also small
wrong-sign (WS) peaking backgrounds in single $\bar D^0$ tags. The
doubly Cabibbo suppressed decays (DCSD) contribute to the WS peaking
background for single $\bar D^0$ tag modes of $\bar D^0\to K^+\pi^-$,
$\bar D^0\to K^+\pi^-\pi^0$ and $\bar D^0\to K^+\pi^-\pi^-\pi^+$.
In addition, the $\bar D^0\to K_S^0K^-\pi^+$ ($K_S^0\to\pi^+\pi^-$),
$\bar D^0\to K_S^0K^-\pi^+\pi^0$ ($K_S^0\to\pi^+\pi^-$) and
$K_S^0K^-\pi^+$ ($K_S^0\to\pi^0\pi^0$) also  make significant
contributions to WS peaking backgrounds for the $\bar D^0\to
K^+\pi^-\pi^-\pi^+$, $\bar D^0\to K^+\pi^-\pi^-\pi^+\pi^0$ and $\bar
D^0\to K^+\pi^-\pi^0\pi^0$ tag modes, respectively. The size of these
WS peaking backgrounds are estimated from Monte Carlo simulation and
then subtracted from the yields obtained from the fits to $M_{\rm BC}$
spectra.

Table~\ref{Tab:d_Ntag} summarizes the single $\bar D^0$ tags.
In the table, the second column gives the
$\Delta E$ requirement on the $Kn\pi$ combination,
the fourth column gives the number of the single $\bar D^0$ tags
in the tag mass region as shown in the third column.

The efficiencies for reconstruction of the single $\bar D^0$ tags
for the five tag modes are obtained by applying the identical analysis
procedure to simulated ``Signal'' Monte Carlo events
mixed with ``Background'' Monte Carlo events. 
The ``Signal'' Monte Carlo events
are generated as
$e^+e^-\to\psi(3770)\to D^0\bar D^0$, where the $\bar D^0$ meson is set to
decay to the tag mode in question and the $D^0$ meson is set to decay
to all possible final states with corresponding branching fractions~\cite{pdg2014}.
The efficiencies for reconstruction of the single $\bar D^0$ tags
are presented in the last column of Table~\ref{Tab:d_Ntag}.

\begin{table*}
\caption{Summary of the single $\bar D^0$ tags
and efficiencies for reconstruction of the single $\bar D^0$ tags,
where $\Delta E$ gives the requirements on the energy difference between the measured $E_{Kn\pi}$
and beam energy $E_{\rm beam}$,
while the $M_{\rm B}$ range defines the signal region of the single $\bar D^0$ tags.
$N_{\rm tag}$ is the number of single $\bar D^0$ tags
and $\epsilon_{\rm tag}$ is the efficiency for reconstruction of the single $\bar D^0$ tags.
}
\label{Tab:d_Ntag}
\begin{ruledtabular}
\begin{tabular}{lcccccc}
 Tag mode                       & $\Delta E$ (GeV) & $M_{\rm BC}$ range (GeV$/c^2$)  & $N_{\rm tag}$ &  $\epsilon_{\rm tag}$ (\%) \\
 \hline
 $K^+\pi^-$                     & $(-0.049, 0.044)$ & $(1.860, 1.875)$ &  $~567083 \pm 848~$  & $70.29 \pm 0.07$   \\
 $K^+\pi^-\pi^0$                & $(-0.071, 0.052)$ & $(1.858, 1.875)$ &  $1094081 \pm 1692$  & $36.80 \pm 0.03$   \\
 $K^+\pi^-\pi^-\pi^+$           & $(-0.043, 0.043)$ & $(1.860, 1.875)$ &  $~700061 \pm 1121$  & $39.57 \pm 0.04$   \\
 $K^+\pi^-\pi^-\pi^+\pi^0$      & $(-0.067, 0.066)$ & $(1.858, 1.875)$ &  $~158367 \pm 749~$  & $15.95 \pm 0.08$   \\
 $K^+\pi^-\pi^0\pi^0$           & $(-0.082, 0.050)$ & $(1.858, 1.875)$ &  $~273725 \pm 2859$  & $15.78 \pm 0.08$   \\
 \hline
 Sum                            &                   &                  &  $2793317 \pm 3684$  \\
\end{tabular}
\end{ruledtabular}
\end{table*}

\subsection{\boldmath Selection of $D^0\rightarrow K^-e^+\nu_e$ and $D^0\rightarrow \pi^-e^+\nu_e$ }
\label{sec:sel_SL}

The $D^0\to K^-e^+\nu _e$ and $D^0\to \pi ^-e^+\nu _e$ event candidates are
selected from the tracks recoiling against the single
$\bar D^0$ tags. To select the $D^0 \to K^-e^+\nu_e$ and
$D^0 \to \pi ^-e^+\nu_e$ events, it is required that there are only
two oppositely charged tracks,
one of which is identified as a positron and
the other as a kaon or a pion.
The combined confidence level $CL_K$ ($CL_{\pi}$) for the
$K$ ($\pi$) hypothesis  is required to be greater than $CL_{\pi}$ ($CL_K$) for kaon (pion) candidates.
For positron identification,
the combined confidence level ($CL_{e}$), calculated for the $e$ hypothesis using
the $dE/dx$, TOF and EMC measurements
(deposited energy and shape of the electromagnetic shower), is required to be greater than
$0.1\%$, and the ratio
$CL_e/(CL_e + CL_{\pi} + CL_K)$ is required
to be greater than $0.8$.
We include the 4-momenta of near-by photons with
the direction of the positron momentum
to partially account for final-state-radiation energy losses (FSR recovery).
In addition, to suppress fake photon background
it is required that
the maximum energy of any unused photon in the recoil system,
$E_{\gamma,\rm max}$, be less than 300~MeV.

Since the neutrino escapes detection,
the kinematic variable
\begin{equation}
    U_{\rm miss}\equiv E_{\rm miss}- |\vec p_{\rm miss}|
\end{equation}
is used to obtain the information about the missing neutrino,
where $E_{\rm miss}$ and $\vec p_{\rm miss}$ are, respectively, the total missing energy
and momentum in the event, computed from
\begin{equation}
 E_{\rm miss} = E_{\rm beam} - E_{h^-} - E_{e^+},
\end{equation}
where
$E_{h^-}$ and $E_{e^+}$ are the measured energies of the hadron and the positron, respectively.
The $\vec p_{\rm miss}$ is calculated by
\begin{equation}
 \vec p_{\rm miss} = \vec p_{D^0} - \vec p_{h^-} - \vec p_{e^+},
\end{equation}
where $\vec p_{D^0}$, $\vec p_{h^-}$ and $\vec p_{e^+}$ are
the momenta of the $D^0$ meson, the hadron and the positron, respectively.
The 3-momentum $\vec p_{D^0}$ of the $D^0$ meson is computed by
\begin{equation}
 \vec p_{D^0} = - \hat p_{\rm tag} \sqrt{E_{\rm beam}^2 - m_{D^0}^2},
\end{equation}
where $\hat p_{\rm tag}$ is the direction of the momentum of the single $\bar D^0$ tag.
If the daughter particles from a semileptonic decay are correctly identified,
$U_{\rm miss}$ is zero, since only one neutrino is missing.

\begin{figure}
\centerline{
\includegraphics[width=0.5\textwidth]{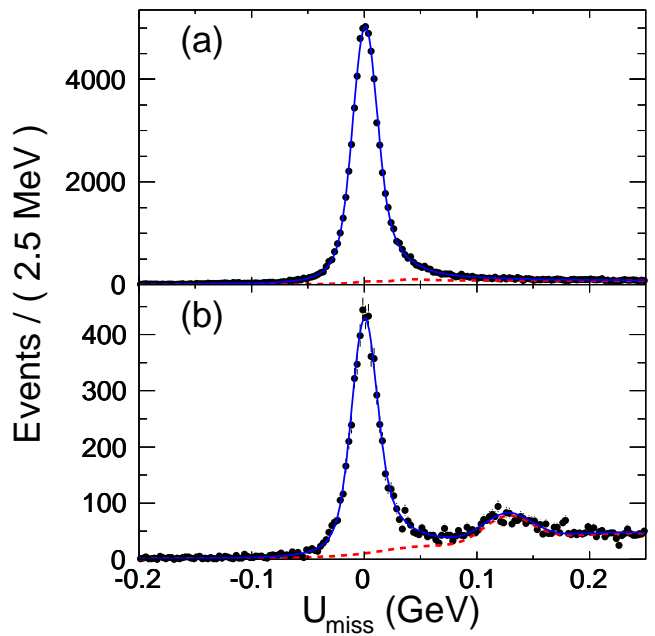}
}
\caption{$U_{\rm miss}$ distributions of events for (a) $\bar D^0$ tags vs. $D^0 \rightarrow K^-e^+\nu_e$,
and for (b) $\bar D^0$ tags vs. $D^0 \rightarrow \pi^-e^+\nu_e$,
where the dots with error bars show the data, the solid lines
show the best fit to the data, and the dashed lines show the background shapes estimated
by analyzing the ``cocktail vs. cocktail $D\bar D$ process'' Monte Carlo events
and the ``non-$D\bar D$ process'' Monte Carlo events (see text for more details).}
\label{Fig:d_Umiss_bc}
\end{figure}

Figures~\ref{Fig:d_Umiss_bc} (a) and (b)
show the $U_{\rm miss}$ distributions for the $D^0 \to K^-e^+\nu_e$ and $D^0 \to \pi^-e^+\nu_e$ candidate events, respectively.
In both cases, most of the events are from the $D^0 \to K^-e^+\nu_e$ and $D^0 \to \pi^-e^+\nu_e$ decays.
Backgrounds from $D\bar D$ processes include mistagged $\bar D^0$ and $D^0$ decays
other than the semileptonic decay in question.
Other backgrounds are from ``non-$D\bar D$ process'' processes.
From the simulated ``cocktail vs. cocktail $D\bar D$ process'' events,
we find that the $D\bar D$ background events are mostly from
$D^0 \to K^- \pi^0 e^+\nu_e$, $D^0 \to K^-\mu^+\nu_{\mu}$
and $D^0 \to \pi^-e^+\nu_e$
selected as $D^0 \to K^-e^+\nu_e$,
and $D^0 \to \pi^- \pi^0 e^+\nu_e$, $D^0 \to K^-e^+\nu_{\mu}$
and $D^0 \to \pi^-\mu^+\nu_{\mu}$
selected as $D^0 \to \pi^-e^+\nu_e$.
Backgrounds from ``non-$D\bar D$'' processes include
the ISR (Initial State Radiation) return to the $\psi(3686)$ and $J/\psi$,
continuum light hadron production,
$\psi(3770) \rightarrow {\rm non}-D\bar D$ decays
and $e^+e^-\rightarrow \tau^+\tau^-$ events.
The levels of these backgrounds events are estimated
by analyzing the corresponding simulated event samples.

Because of  ISR and FSR (Final State Radiation), the signal $U_{\rm miss}$ distributions are not Gaussian;
instead, the $U_{\rm miss}$ distributions have Gaussian cores with
long tails at both the lower and the higher sides of the distributions.
To obtain the numbers of the signal events for these two semileptonic decays,
we fit these distributions with an empirical function that includes these tails.

We use the same probability density function as CLEO~\cite{cleo-c_Phys_Rev_D79_052010_y2009}
for $U_{\rm miss}$,
\begin{equation}
f(x) = \left\{
 \begin{array}{ll}
	a_{1}\left(\frac{n_{1}}{\alpha_{1}}-\alpha_{1}+x\right )^{-n_{1}}
	& {\rm if~} x \geq \alpha_{1} \\
	\exp(-x^2/2)
	& {\rm if~} -\alpha_{2}\leq x <\alpha_{1}, \\
	a_{2}\left(\frac{n_{2}}{\alpha_{2}}-\alpha_{2}-x\right )^{-n_{2}}
	& {\rm if~} x <-\alpha_{2}
 \end{array} \right.
\label{f_umiss}
\end{equation}
where $x\equiv (U_{\rm miss}-m)/\sigma$, $m$ and $\sigma$ are the mean value and
standard deviation of the Gaussian distribution, respectively.
In Eq.~(\ref{f_umiss}), $a_{1}\equiv(n_{1}/\alpha_{1})^{n_{1}}e^{-\alpha^{2}_{1}/2}$,
and $a_{2}\equiv(n_{2}/\alpha_{2})^{n_{2}}e^{-\alpha^{2}_{2}/2}$,
where $\alpha_1$, $\alpha_2$, $n_1$ and $n_2$ are
parameters describing the tails of the signal function,
determined from fits to the simulated $U_{\rm miss}$ distributions of
signal Monte Carlo events.

To account for differences between data and Monte Carlo,
we fit the data using the Monte Carlo determined $f(x)$
distribution convolved with a Gaussian function with free mean and width.
The background function
is formed from histograms of $U_{\rm miss}$ distributions for background events
from the ``cocktail vs. cocktail'' $D\bar D$
and ``non-$D\bar D$'' simulated event samples.
The normalizations of the signal and background are free parameters in the fits to
the data.

The results of the fits to the two $U_{\rm miss}$ distributions are shown in
Figs.~\ref{Fig:d_Umiss_bc} (a) and (b);
the fitted yields of signal events are
\begin{equation}
N_{\rm observed} (D^0\to K^-e^+\nu_e) = 70727.0 \pm 278.3
\end{equation}
and
\begin{equation}
N_{\rm observed} (D^0\to \pi^-e^+\nu_e) = 6297.1 \pm 86.8.
\end{equation}
In Fig.~\ref{Fig:d_Umiss_bc} (a) and (b),
the solid lines show the best fits to the data,
while the dashed lines show the background.

\begin{figure*}
\centerline{
\includegraphics[width=0.75\textwidth]{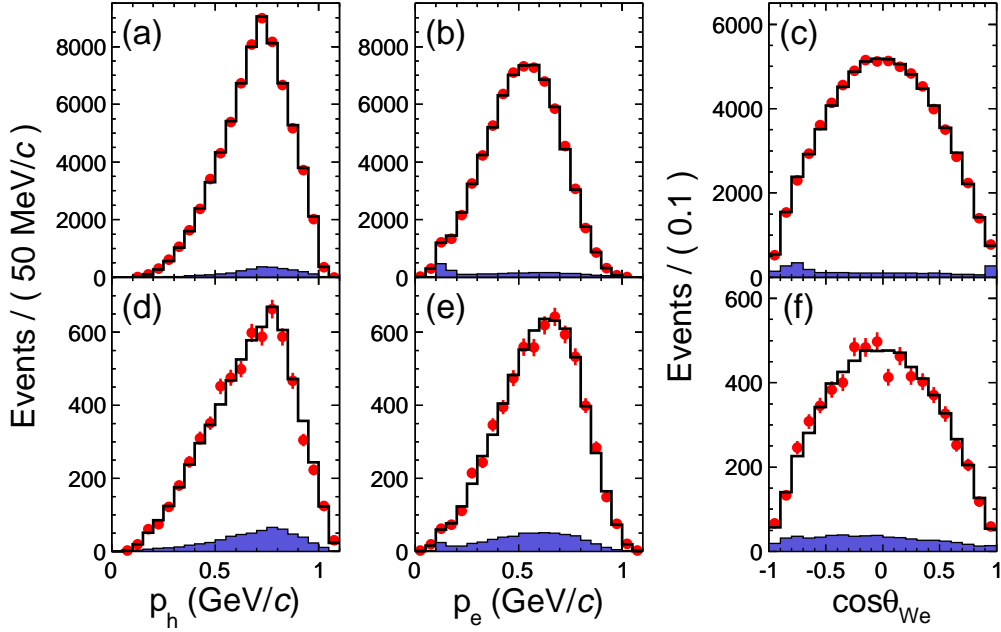}
}
\caption{
Distributions of particle momenta and $\cos\theta_{We}$ from
$D^0 \rightarrow K^-e^+\nu_e$ and $D^0 \rightarrow \pi^-e^+\nu_e$ semileptonic decays,
where (a) and (b) are the momenta of kaon and positron from $D^0 \rightarrow K^-e^+\nu_e$, respectively,
(d) and (e) are the momenta of pion and positron from $D^0 \rightarrow \pi^-e^+\nu_e$, respectively;
(c) and (f) are the distributions of $\cos\theta_{We}$
for $D^0\to K^-e^+\nu_e$ and $D^0\to \pi^-e^+\nu_e$, respectively;
these events satisfy $-0.06<U_{\rm miss}<0.06~\rm GeV$.
The  solid histograms are Monte Carlo simulated signal plus background;
the shaded histograms are Monte Carlo simulated background only.
}
\label{fig:Cmp_p_cos}
\end{figure*}

To gain confidence in the quality of the Monte Carlo simulation,
we examine the momentum distributions
of the kaon, the pion and the positron as well as $\cos\theta_{We}$ from the semileptonic decays of
$D^0 \rightarrow K^-e^+\nu_e$ and $D^0 \rightarrow \pi^-e^+\nu_e$,
where $\theta_{We}$ is the angle
between the direction of the virtual $W^+$ boson in the $D^0$ rest frame
and the three-momentum of the positron in the $W^+$ rest frame.
These distributions are shown in Figs.~\ref{fig:Cmp_p_cos} (a)-(f), respectively,
where the dots with error bars are for the data,
the  solid histograms are for the full Monte Carlo simulation
and the shaded histograms show the Monte Carlo simulated backgrounds only.

\section{Measurements of Absolute Decay Branching Fractions}
\label{sec:bf}

\subsection{Efficiency for reconstruction of semileptonic decays}

To determine the efficiency $\epsilon(D^0\rightarrow h^-e^+\nu_e)$
for reconstruction of each of the two semileptonic decays for each single tag mode,
``Signal'' Monte Carlo event samples of
$\psi(3770)\to D^0\bar D^0$ decays, where the $D^0$ meson is set to decay to the
$h^-e^+\nu_e$ final state in question and the $\bar D^0$ meson is set to decay to
each of the five single $\bar D^0$ tag modes,
are generated and simulated with the BESIII software package.
By subjecting these simulated events to
the same requirements that are applied to the data
we obtain the reconstruction efficiencies $\epsilon_{{\rm tag}, D^0 \to h^-e^+\nu_e}$
for simultaneously finding the $D^0$ meson semileptonic decay
and the single $\bar D^0$ tag in the same event; these are
given in Tab.~\ref{tab_eff_Kenu_pienu}.
\begin{table*}
 \caption{Double tag efficiencies for reconstruction of ``$\bar D^0_{\rm tag}$ vs. $D^0 \to h^-e^+ \nu_e$''
and overall efficiencies for reconstruction of $D^0 \to h^-e^+ \nu_e$ in the recoil side of $\bar D^0$ tags.}
 \label{tab_eff_Kenu_pienu}
 \begin{ruledtabular}
 \begin{tabular}{lcccc}
  Tag mode
  & $\epsilon_{{\rm tag}, D^0\to K^-e^+\nu_e}$
  & $\epsilon_{{\rm tag}, D^0\to \pi^-e^+\nu_e}$
  & $\epsilon_{\rm MC}(D^0\to K^-e^+\nu_e)$
  &  $\epsilon_{\rm MC}(D^0\to \pi^-e^+\nu_e)$  \\
  \hline
  $K^{+}\pi^{-}$                      & $0.4566 \pm 0.0014$  &  $0.4995 \pm 0.0014$ & $0.6496 \pm 0.0021$ & $0.7106 \pm 0.0021$ \\
  $K^{+}\pi^{-}\pi^{0}$               & $0.2685 \pm 0.0006$  &  $0.2927 \pm 0.0007$ & $0.7296 \pm 0.0017$ & $0.7954 \pm 0.0020$ \\
  $K^{+}\pi^{-}\pi^{-}\pi^{+}$        & $0.2666 \pm 0.0008$  &  $0.2897 \pm 0.0008$ & $0.6737 \pm 0.0021$ & $0.7321 \pm 0.0022$ \\
  $K^{+}\pi^{-}\pi^{-}\pi^{+}\pi^{0}$ & $0.1260 \pm 0.0008$  &  $0.1363 \pm 0.0008$ & $0.7900 \pm 0.0064$ & $0.8545 \pm 0.0066$ \\
  $K^{+}\pi^{-}\pi^{0}\pi^{0}$        & $0.1331 \pm 0.0007$  &  $0.1467 \pm 0.0007$ & $0.8435 \pm 0.0062$ & $0.9297 \pm 0.0065$ \\
  \hline
  Average                             &                      &                      & $0.7140 \pm 0.0012$ & $0.7788 \pm 0.0013$ \\
 \end{tabular}
 \end{ruledtabular}
\end{table*}

Due to their low multiplicity, it is usually easier to reconstruct
$\bar D^0$ tags in semileptonic events than in typical $D^0\bar D^0$
events (tag bias). In addition, the size of the tag bias is correlated
with the multiplicity of the tag mode. In consequence the overall
efficiencies shown in Tab.~\ref{tab_eff_Kenu_pienu} vary greatly from
the $\bar D^0\to K^-\pi^+$ mode to the $\bar D^0\to K^-\pi^+\pi^+\pi^-$ and
$\bar D^0\to K^-\pi^+\pi^0\pi^0$ modes.

The last row in Tab.~\ref{tab_eff_Kenu_pienu} gives the overall efficiency
which is obtained by weighting the individual efficiencies for each of the five single $\bar D^0$ tags
by the corresponding yield shown in Tab.~\ref{Tab:d_Ntag}.

There are small differences in efficiencies for finding a charged particle
and for identifying the type of the charged particle
between the data and Monte Carlo events
that are discussed below in Section~\ref{sec:syst}.
To take these differences into account, the overall efficiencies
$\epsilon_{\rm MC}(D^0\to K^-e^+\nu_e)$ and $\epsilon_{\rm MC}(D^0\to \pi^-e^+\nu_e)$
are corrected by the multiplicative factors of
\begin{equation}
 f_{\rm corr}^{\rm trk+PID} =
 \left\{
 \begin{array}{ll}
         1.0118 & {\rm for}~D^0\to K^-e^+\nu_e, \\
         0.9814 & {\rm for}~D^0\to \pi^-e^+\nu_e.
 \end{array}
 \right .
\label{f_crr_trk_pid}
\end{equation}
After making these corrections, we obtain the ``true'' overall efficiencies for
reconstruction of these two semileptonic decays,
\begin{equation}
    \epsilon(D^0\to K^-e^+\nu_e) = 0.7224\pm0.0012,
\end{equation}
and
\begin{equation}
    \epsilon(D^0\to \pi^-e^+\nu_e) = 0.7643\pm 0.0013.
\end{equation}

\subsection{Decay branching fraction}

   Inserting the number of the single $\bar D^0$ tags,
the numbers of the signal events for these two $D^0$ semileptonic decays
observed in the recoil of the single $\bar D^0$ tags
together with corresponding efficiency into Eq.(\ref{bf_fotmula}),
we obtain the absolute decay branching fractions
\begin{equation}
 \mathcal B(D^0 \to K^-e^+\nu_e) = (3.505\pm 0.014 \pm 0.033)\%
\label{eq_bf_kev}
\end{equation}
and
\begin{equation}
 \mathcal B(D^0 \to \pi^- e^+\nu_e)=(0.295\pm 0.004\pm 0.003)\%,
\label{eq_bf_piev}
\end{equation}
where the first errors are statistical and the second systematic.
The sources of systematic uncertainties in the measured decay branching fractions
are discussed in the next subsection.

\subsection{Systematic uncertainties in measured branching fractions}
\label{sec:syst}

Table~\ref{systematic_err_soures} lists the sources of the systematic uncertainties
in the measured semileptonic branching fractions.
We discuss each of these sources in the following.

\begin{table}[!hbp]
\caption{Sources of the systematic uncertainties
in the measured branching fractions for
$D^0 \rightarrow K^-e^+\nu_e$ and $D^0 \rightarrow \pi^-e^+\nu_e$.}
\label{systematic_err_soures}
\begin{ruledtabular}
\begin{tabular}{lcc}
& \multicolumn{2}{c} {Systematic uncertainty (\%)} \\
Source  & $K^-e^+\nu_e$  & $\pi^-e^+\nu_e$  \\
\hline
Number of $\bar D^0$ tags       & 0.50  &  0.50     \\
Tracking for $e^+$              & 0.19  &  0.15     \\
Tracking for $K^-$              & 0.42  &  ---      \\
Tracking for $\pi^-$            & ---   &  0.28     \\
PID for  $e^+$                  & 0.16  &  0.14     \\
PID for  $K^-$                  & 0.10  &  ---      \\
PID for  $\pi^-$                & ---   &  0.19     \\
$E_{\gamma,{\rm max}}$ cut      & 0.10  &  0.10     \\
Fit to $U_{\rm miss}$           & 0.48  &  0.50     \\
Form factor structure           & 0.10  &  0.10     \\
FSR recovery                    & 0.30  &  0.30     \\
Finite MC statistics          & 0.17  &  0.17     \\
Single tag cancelation          & 0.12  &  0.12     \\
\hline
Total                           & 0.94  &  0.90     \\
\end{tabular}
\end{ruledtabular}
\end{table}

\subsubsection{Uncertainty in number of $\bar D^0$ tags}

To estimate the uncertainty in the number of single $\bar D^0$ tags,
we repeat the fits to the $M_{\rm BC}$ distributions by varying the
bin size, fit range and background functions. We also investigate the
contribution arising from  possible differences in the $\pi^0$ fake
rates between data and Monte Carlo simulation. Finally, we assign a
systematic uncertainty of 0.5\% to the number of $\bar D^0$ tags.

\subsubsection{Uncertainty in tracking efficiency}

The uncertainties for finding a charged track
are estimated by comparing the efficiencies for reconstructing
 the positron, kaon
and pion in data and Monte Carlo events.

Using radiative Bhabha scattering events selected from the data
and simulated radiative Bhabha scattering events,
we measure the difference in efficiencies for finding a positron
between data and simulation.
Considering both the $\cos\theta$, where $\theta$ is the polar angle of the positron,
and momentum distributions of the positrons,
we obtain two-dimensional weighted-average efficiency differences
($\epsilon_{\rm data}/\epsilon_{\rm MC}-1$)
of $(0.22\pm 0.19)\%$ and $(0.11\pm 0.15)\%$.
These translate uncertainties on the decay branching fractions of
$0.19\%$ and $0.15\%$ for $D^0 \to K^-e^+\nu_e$
and $D^0 \to \pi^- e^+\nu_e$ decays, respectively.

The efficiencies for finding a charged kaon and a charged pion are determined
by analyzing doubly tagged $D\bar D$ decay events.
In the selection of the doubly tagged $D\bar D$ decay events,
we exclude one charged kaon or one charged pion track
and examine the variable $M^2_{{\rm miss}~K~{\rm or}~\pi}$,
defined as
the difference between the missing energy squared $E^2_{\rm miss}$ and the
missing momentum squared $p^2_{\rm miss}$ of the selected $D\bar D$ decay events.
By analyzing these $M^2_{{\rm miss}~K~{\rm or}~\pi}$ variables
for both the data and the simulated ``cocktail vs. cocktail $D\bar D$ process'' Monte Carlo events,
we find the differences in efficiencies for reconstructing a charged kaon or
a charged pion between the data and the Monte Carlo events
as a function of the charged particle momentum.
Considering the momentum distributions of the kaon and pion from
these two semileptonic decays,
we obtain the magnitudes of systematic differences and their uncertainties
of the track reconstruction efficiencies.
The level of uncertainties in the corrections
for these differences in measurements of the decay branching fractions
and partial decay rates (see Section \ref{sec:dr}) are $0.42\%$ and $0.28\%$
for charged kaons and pions, respectively.

\subsubsection{Uncertainty in particle identification}

The differences in efficiencies for identifying a positron between the data and the Monte Carlo samples
depend not only on the momentum of the positron, but also on $\cos\theta$.
Considering both of these for our signal positrons,
we obtain a weighted-average difference in efficiency
for identifying the positron from the two semileptonic decays.
After making correction for these differences in efficiencies for
identifying the positrons,
we obtain a systematic uncertainty of $0.16\%$ ($0.14\%$)
on the $K^-(\pi^-)e^+\nu_e$  mode from this source.

The systematic uncertainties
associated with the efficiencies for identifying a charged kaon and a charged pion
are estimated using the missing mass square techniques discussed above.
Taking into account the momentum distributions of the charged particles from the two semileptonic modes,
we correct for the momentum-weighted efficiency differences for
identifying the kaon and the pion, and we assign systematic
uncertainties of $0.10\%$ and $0.19\%$ for charged kaons and pions,
respectively.

\subsubsection{Uncertainty in $E_{\gamma, \rm max}$ cut}

The uncertainty associated with the $E_{\gamma,\rm max}$ requirement on the events is estimated by
analyzing doubly tagged $D\bar D$ events with hadronic decay modes.
With these events,
we examine the fake photons from the EMC measurements.
By analyzing these selected samples from both the data and the simulated Monte Carlo events,
we find that the magnitude of difference in the number of fake photons
between the data and the Monte Carlo events is $0.10\%$,
which is set as the systematic uncertainty due to this source.

\subsubsection{Uncertainty in fit to $U_{\rm miss}$ distribution}
\label{sec:syst_umiss}

To estimate the systematic uncertainty in the numbers of signal events
due to the fit to the $U_{\rm miss}$ distribution,
we vary the bin size and the tail parameters of the signal function.
We then repeat the fits to the $U_{\rm miss}$ distributions, and combine
the changes in the yields in quadrature to obtain the systematic uncertainty.
Since the background function is formed from many background modes
with fixed relative normalizations, we also vary the relative
contributions of several of the largest background modes based on the
uncertainties in their branching fractions and the uncertainties in
the rates of misidentifying a hadron (muon) as an electron.
Finally we find that the relative sizes of this systematic uncertainty
are 0.48\% and 0.50\% for $D^0\to K^-e^+\nu_e$ and
$D^0\to\pi^-e^+\nu_e$, respectively.

\subsubsection{Uncertainty in form factors}

In order to estimate the systematic uncertainty associated with
the form factor used to generate signal events in the Monte Carlo simulation,
we re-weight the signal Monte Carlo events so that their $q^2$ distributions
match the measured spectra. We then re-measure the branching fraction
(partial decay rates in different $q^2$ bins)
with the new weighted efficiency (efficiency matrix).
The maximum relative changes in branching fraction
(partial decay rates in different $q^2$ bins) is $0.05\%$.
To be conservative, we assign a relative systematic uncertainty of $0.10\%$
to the branching fraction measurements for
$D^0\to K^-e^+\nu_e$ and $D^0\to\pi^-e^+\nu_e$ decays.

\subsubsection{Uncertainty in FSR recovery}

The difference between the measured branching fraction obtained with
the FSR recovery of the positron momentum and the one obtained without
the FSR recovery is assigned as the most conservative systematic
uncertainty due to FSR recovery. We find the magnitude of this
difference to be $0.30\%$ for both $D^0\to K^-e^+\nu_e$
and $D^0\to \pi^-e^+\nu_e$ decays.

\subsubsection{Uncertainty due to finite Monte Carlo statistics}

The uncertainties associated with the
finite Monte Carlo statistics
are $0.17\%$ for both $D^0 \rightarrow K^-e^+\nu_e$ and
$D^0 \rightarrow \pi^-e^+\nu_e$.

\subsubsection{Uncertainty due to single tag cancelation}
Most of the systematic uncertainties arising from the selection of
single $\bar D^0$ tags are canceled due to the double tag technique.
The un-canceled systematic error of MDC tracking, particle identification
and $\pi^0$ selection in single tag selection is estimated by
$\left(\sum_{\rm tag} (\epsilon^\prime_{\rm tag}/
\epsilon_{\rm tag}-1 )\times 0.25\delta_{\rm tag}
\times N_{\rm tag}\right)/\left(\sum_{\rm tag} N_{\rm tag}\right)$,
where
$\epsilon^\prime_{\rm tag}$ and $\epsilon_{\rm tag}$ are the efficiencies
of reconstructing single $\bar D^0$ tags obtained by analyzing the Monte Carlo
events of $\bar D^0\to {\rm tag}$ vs. $D^0\to h^-e^+\nu_e$  and
$\bar D^0\to {\rm tag}$ vs. $D^0\to anything$ after mixing all the simulated
backgrounds, respectively;
$N_{\rm tag}$ is the number of single $\bar D^0$ tags reconstructed
in data;
$\delta_{\rm tag}$ is the total systematic error of MDC tracking, particle
identification and $\pi^0$ selection in single tag selection.
Since no efficiency correction is made in the single tag selection,
the uncertainty in MDC tracking (or particle identification) for charged
kaon or pion is taken to be 1.0\% per track, and the uncertainty in
$\pi^0$ selection is taken to be 2.0\% per $\pi^0$.
For each single $\bar D^0$ tag mode, the uncertainty in MDC
tracking, particle identification or $\pi^0$ selection are added linearly
separately, and then they are added in quadrature to obtain the total
systematic error in the single $\bar D^0$ tag selection.
Finally,
we assign a systematic uncertainty of 0.12\%
for the branching fraction measurements.

\subsection{Comparison with other measurements}

A comparison of our measured branching fractions
for $D^0 \rightarrow K^-e^+\nu_e$ and
$D^0 \rightarrow \pi^-e^+\nu_e$ decays with those
previously measured by the MARK-III~\cite{mark-iii},
CLEO~\cite{cleo},
BES-II~\cite{bes2_plb597_p39_2004},
CLEO~\cite{cleo-c_prl95_181802_y2005,
cleo-c_Phys_Rev_D79_052010_y2009, cleo-c_Phys_Rev_D80_032005_y2009}
(at the CLEO-c experiment) and \babar~\cite{BaBar_Phys_Rev_D76_p052005_y2007,BaBar_D0topienu} Collaborations
as well as the world average given by the PDG~\cite{pdg2014}
is given in Table~\ref{tab_cmp_bf}.
Our measured branching fractions for these two decays are in excellent agreement
with the experimental results obtained by other experiments, but are more precise.
In the table, we also compare our  branching fraction measurements
to theoretical predictions for these two semileptonic decays.
The precision of our measured branching fractions are much higher
than those of the LQCD~\cite{lqcd_prl_94_011601, LQCD_2},
the QCD sum rule~\cite{QCDSR} and the LCSR~\cite{LCSR_2} predictions.
\begin{table*}
\caption{Comparison of  the measured $\mathcal B(D^0\to K^-e^+\nu_e)$ and $\mathcal B(D^0\to \pi^-e^+\nu_e)$
values with those measured by other experiments and theoretical predictions based on QCD  and the world-average value of $\tau_{D^0}$.
}
\label{tab_cmp_bf}
\begin{ruledtabular}
\begin{tabular}{lcc}
Experiment/Theory & $\mathcal B(D^0\to K^-e^+\nu_e)$ (\%) & $\mathcal B(D^0\to \pi^-e^+\nu_e)$ (\%) \\
\hline
PDG2014~\cite{pdg2014}               & $3.55 \pm 0.05$            & $0.289 \pm 0.008$             \\
MARK-III~\cite{mark-iii}             & $3.4\pm 0.5\pm 0.4$      & $0.39^{+0.23}_{-0.11} \pm 0.04$ \\
CLEO~\cite{cleo}                     & $3.82\pm 0.11\pm 0.25$   &                                 \\
BES-II~\cite{bes2_plb597_p39_2004}   & $3.82 \pm 0.40 \pm 0.27$ & $0.33\pm 0.13\pm 0.03$          \\
CLEO-c~\cite{cleo-c_Phys_Rev_D80_032005_y2009} & $3.50\pm 0.03\pm 0.04$   & $0.288 \pm 0.008\pm 0.003$     \\
Belle~\cite{bell_Phys_RevLett_97_p061804_y2006}   & $3.45\pm 0.07 \pm 0.20$ & $0.255\pm 0.019\pm 0.016$  \\
\babar~\cite{BaBar_Phys_Rev_D76_p052005_y2007,BaBar_D0topienu} & $3.522\pm 0.027 \pm 0.045\pm 0.065$  &   $0.2770\pm0.0068\pm0.0092\pm0.0037$               \\
BESIII (this experiment)              & $3.505\pm 0.014 \pm 0.033$ & $0.295\pm 0.004\pm 0.003$  \\
\hline
LQCD ~\cite{lqcd_prl_94_011601}                  & $3.77 \pm 0.29\pm 0.74$    & $0.316 \pm 0.025\pm 0.070$    \\
LQCD ~\cite{LQCD_2}                  & $2.99 \pm 0.45\pm 0.74$    & $0.24 \pm 0.06$               \\
QCD SR~\cite{QCDSR}                  & $2.7 \pm 0.6$              &                 \\
LCSR~\cite{LCSR_2}                   & $3.9 \pm 1.2 $            & $0.30 \pm 0.09 $               \\
\end{tabular}
\end{ruledtabular}
\end{table*}

\section{Differential decay rates}
\label{sec:dr}

The differential decay rate $d\Gamma/d q^2$ for $D^0\to K^-(\pi^-)e^+\nu_e$
is given by Eq.~(\ref{eq_dGamma_dq2}).
The form factor $f_+^{K(\pi)}(q^2)$ can be extracted from
measurements of $d\Gamma/dq^2$. Such measurements are obtained from
the event rates in bins of $q^2$ ranging from $q^2_i - 0.5 \Delta q^2$
to $q^2_i + 0.5 \Delta q^2$, where $\Delta q^2$ is the bin width
and $i$ is the bin number.

\subsection{Measurement of differential decay rates}

The $q^{2}$ value is given by
\begin{equation}
q^2 = (E_{e}+E_{\nu})^2 - (\vec p_{e} + \vec p_{\nu})^2,
\end{equation}
where $E_e$ and $\vec p_e$ are the measured energy and momentum of the positron,
$E_{\nu}$ and $\vec p_{\nu}$ are the energy and momentum of the missing neutrino:
\begin{equation}
 E_{\nu} = E_{\rm miss},
\end{equation}
\begin{equation}
 \vec p_{\nu} = E_{\rm miss}\hat p_{\rm miss}.
\end{equation}

For the $D^0 \rightarrow K^-e^+\nu_e$ differential rate,
we divide the candidates for the decays into 18 $q^{2}$ bins.
For the $D^{0}\to \pi^{-}e^{+}\nu_{e}$ mode,
which has fewer events,
we use 14 $q^{2}$ bins.
The first columns of Tables~\ref{tab_Nobs_q2_K} and \ref{tab_Nobs_q2_pi}
give the range of each $q^2$ bin
for $D^{0}\to K^{-}e^{+}\nu_{e}$ and $D^{0}\to \pi^{-}e^{+}\nu_{e}$, respectively.

\begin{table}[!hbp]
\centering
\caption{
Summary of the range of each $q^2$ bin,
the number of the observed events $N_{\rm observed}$,
the number of produced events $N_{\rm produced}$,
and the partial decay rate $\Delta\Gamma$ in each $q^2$ bin
for $D^0\to K^-e^+\nu_e$ decays.
}
\label{tab_Nobs_q2_K}
\begin{ruledtabular}
\begin{tabular}{lccc}
$q^2$ (GeV$^2/c^4)$ & $N_{\rm observed}$   & $N_{\rm produced}$ & $\Delta\Gamma$ (ns$^{-1}$)   \\
\hline
(0.0, 0.1)             &  $7876.1\pm94.2$   & $10094.9\pm132.3$ & $ 8.812\pm0.116$ \\
(0.1, 0.2)             &  $7504.3\pm90.5$   & $10015.4\pm140.8$ & $ 8.743\pm0.123$ \\
(0.2, 0.3)             &  $6940.5\pm87.2$   & $~9502.6\pm142.0$ & $ 8.295\pm0.124$ \\
(0.3, 0.4)             &  $6376.0\pm83.4$   & $~8667.9\pm138.6$ & $ 7.567\pm0.121$ \\
(0.4, 0.5)             &  $6139.8\pm81.9$   & $~8575.9\pm137.7$ & $ 7.486\pm0.120$ \\
(0.5, 0.6)             &  $5460.5\pm77.1$   & $~7384.0\pm128.1$ & $ 6.446\pm0.112$ \\
(0.6, 0.7)             &  $5120.3\pm74.7$   & $~7101.8\pm125.8$ & $ 6.200\pm0.110$ \\
(0.7, 0.8)             &  $4545.5\pm70.5$   & $~6322.2\pm120.2$ & $ 5.519\pm0.105$ \\
(0.8, 0.9)             &  $4159.4\pm67.1$   & $~5760.3\pm113.3$ & $ 5.028\pm0.099$ \\
(0.9, 1.0)             &  $3680.7\pm63.2$   & $~5183.5\pm107.6$ & $ 4.525\pm0.094$ \\
(1.0, 1.1)             &  $3199.6\pm58.9$   & $~4550.0\pm100.2$ & $ 3.972\pm0.087$ \\
(1.1, 1.2)             &  $2637.1\pm53.5$   & $~3810.2\pm92.4~$ & $ 3.326\pm0.081$ \\
(1.2, 1.3)             &  $2239.1\pm49.4$   & $~3239.1\pm84.3~$ & $ 2.828\pm0.074$ \\
(1.3, 1.4)             &  $1752.1\pm43.9$   & $~2621.2\pm77.3~$ & $ 2.288\pm0.067$ \\
(1.4, 1.5)             &  $1301.0\pm37.7$   & $~1989.4\pm67.4~$ & $ 1.737\pm0.059$ \\
(1.5, 1.6)             &  $ 927.5\pm32.0$   & $~1505.1\pm59.0~$ & $ 1.314\pm0.052$ \\
(1.6, 1.7)             &  $ 541.3\pm24.6$   & $~~983.4\pm50.3~$ & $ 0.858\pm0.044$ \\
(1.7, $q^2_{\rm max}$) &  $ 188.2\pm15.1$   & $~~434.2\pm39.6~$ & $ 0.379\pm0.035$ \\
\end{tabular}
\end{ruledtabular}
\label{Tab:PDR_KeV}
\end{table}

\begin{table}[!hbp]
\centering
\caption{
Summary of the range of each $q^2$ bin,
the number of the observed events $N_{\rm observed}$,
the number of produced events $N_{\rm produced}$,
and the partial decay rate $\Delta\Gamma$ in each $q^2$ bin
for $D^0\to \pi^-e^+\nu_e$ decays.
}
\label{tab_Nobs_q2_pi}
\begin{ruledtabular}
\begin{tabular}{lccc}
 $q^2$ (GeV$^2/c^4$) & $N_{\rm observed}$   & $N_{\rm produced}$ & $\Delta\Gamma$ (ns$^{-1}$)    \\
 \hline
 (0.0, 0.2)             &  $ 814.4\pm30.9$  &  $1066.9\pm43.2$ & $0.931\pm0.038$ \\
 (0.2, 0.4)             &  $ 697.2\pm28.7$  &  $~935.1\pm42.8$ & $0.816\pm0.037$ \\
 (0.4, 0.6)             &  $ 634.6\pm27.7$  &  $~836.6\pm41.3$ & $0.730\pm0.036$ \\
 (0.6, 0.8)             &  $ 654.6\pm27.8$  &  $~850.1\pm40.6$ & $0.742\pm0.035$ \\
 (0.8, 1.0)             &  $ 643.2\pm27.3$  &  $~840.2\pm39.9$ & $0.733\pm0.035$ \\
 (1.0, 1.2)             &  $ 578.6\pm26.3$  &  $~744.6\pm37.7$ & $0.650\pm0.033$ \\
 (1.2, 1.4)             &  $ 509.9\pm24.7$  &  $~651.1\pm35.1$ & $0.568\pm0.031$ \\
 (1.4, 1.6)             &  $ 438.6\pm23.2$  &  $~551.6\pm32.8$ & $0.481\pm0.029$ \\
 (1.6, 1.8)             &  $ 412.6\pm22.3$  &  $~534.7\pm31.7$ & $0.467\pm0.028$ \\
 (1.8, 2.0)             &  $ 320.9\pm19.8$  &  $~420.6\pm28.6$ & $0.367\pm0.025$ \\
 (2.0, 2.2)             &  $ 245.8\pm17.0$  &  $~324.0\pm24.7$ & $0.283\pm0.022$ \\
 (2.2, 2.4)             &  $ 165.4\pm14.1$  &  $~229.9\pm21.7$ & $0.201\pm0.019$ \\
 (2.4, 2.6)             &  $  93.6\pm10.7$  &  $~129.2\pm16.7$ & $0.113\pm0.015$ \\
 (2.6, $q^2_{\rm max}$)       &  $  75.8\pm10.0$  &  $~107.2\pm15.0$ & $0.094\pm0.013$ \\
\end{tabular}
\end{ruledtabular}
\label{Tab:PDR_PieV}
\end{table}

\begin{figure*}
\centerline{
\includegraphics[width=0.75\textwidth]{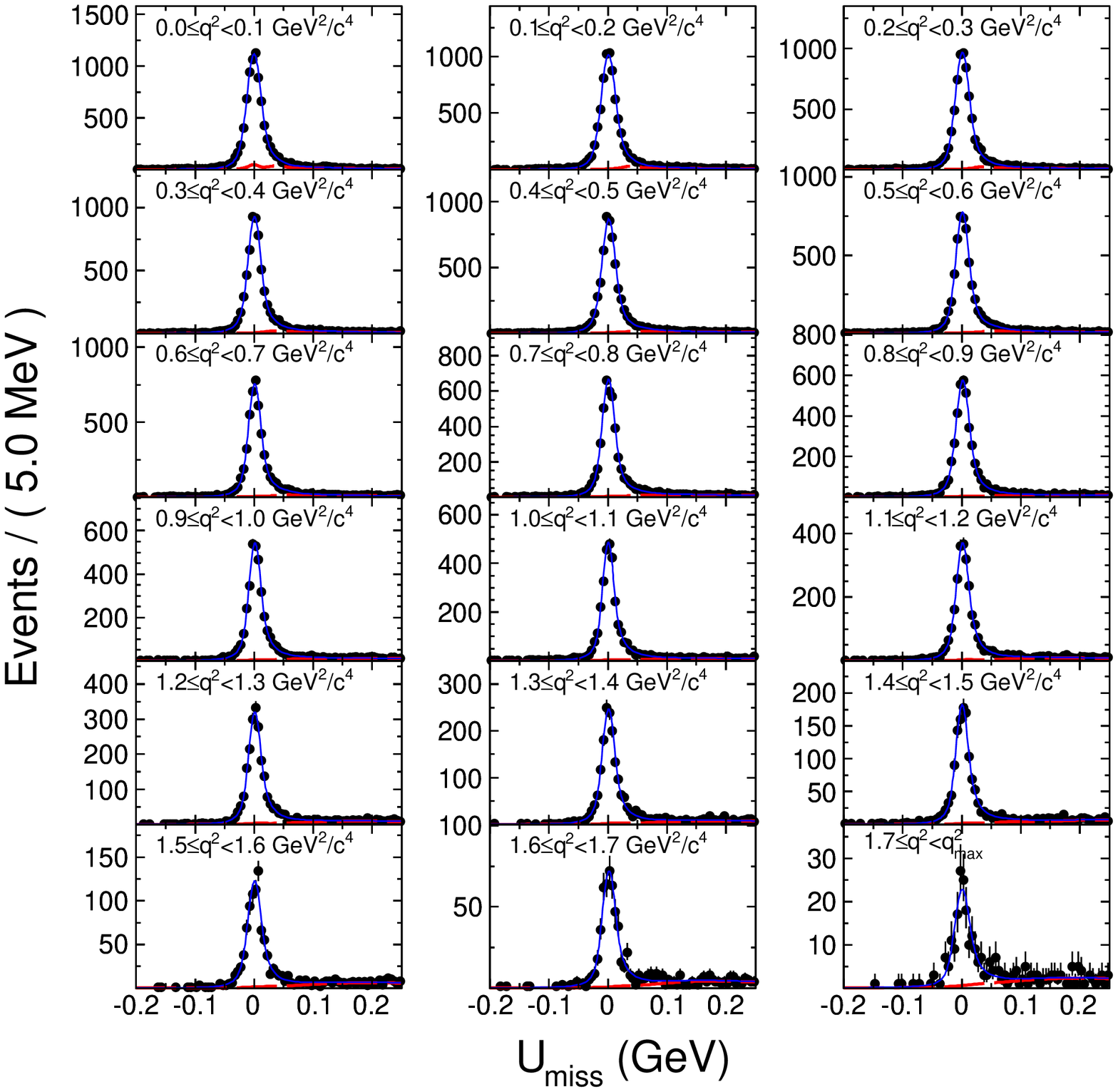}
}
\caption{Distributions of $U_{\rm miss}$ for $\bar D^0$ tags vs. $D^0 \rightarrow K^-e^+\nu_e$
with the squared 4-momentum transfer $q^2$ filled in different $q^2$ bins.
The dots with error bars show the data, the blue solid lines show the best fits to the data,
while the red dashed lines show the background shapes.
}
\label{fig_umiss_Kenu_18_q2_bins}
\end{figure*}

\begin{figure*}
\centerline{
\includegraphics[width=0.75\textwidth]{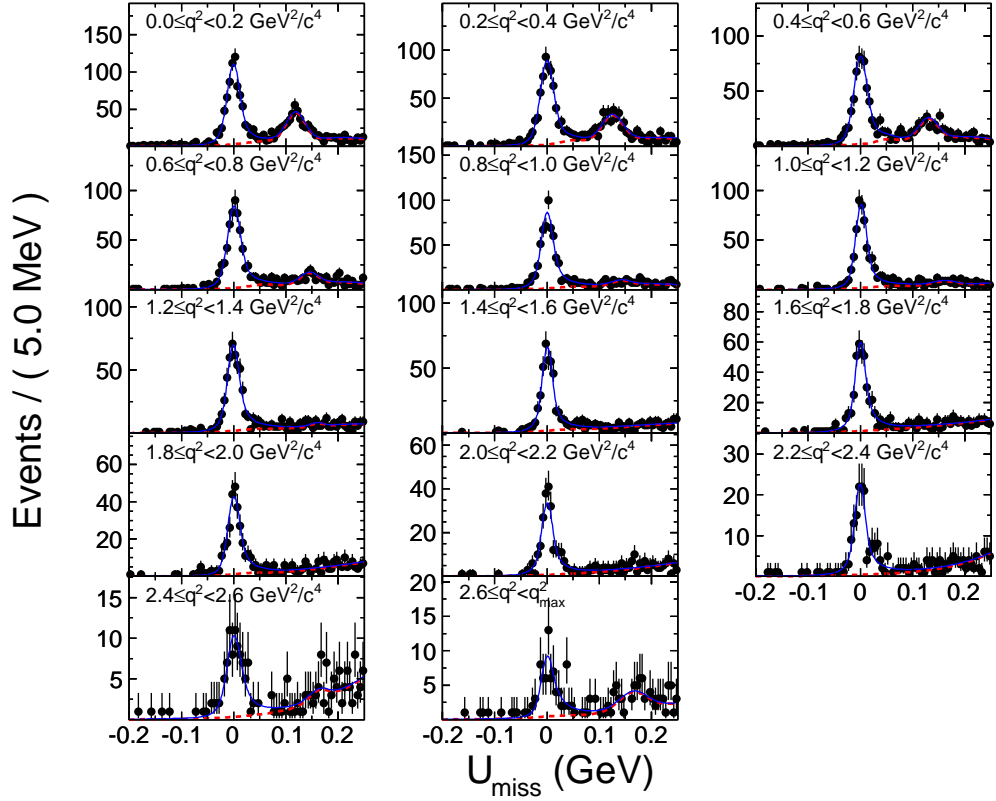}
}
\caption{Distributions of $U_{\rm miss}$ for $\bar D^0$ tags vs. $D^0 \rightarrow \pi^-e^+\nu_e$
with the squared 4-momentum transfer $q^2$ filled in different $q^2$ bins.
The dots with error bars show the data, the blue solid lines show the best fits to the data,
while the red dashed lines show the background shapes.
}
\label{fig_umiss_pienu_14_q2_bins}
\end{figure*}

The points with error bars in
Figs.~\ref{fig_umiss_Kenu_18_q2_bins} and \ref{fig_umiss_pienu_14_q2_bins} show
the $U_{\rm miss}$ distributions for the $D^0\rightarrow K^-e^+\nu_e$
and $D^0\rightarrow \pi^-e^+\nu_e$ decays
for each $q^2$ bin, respectively.
Fits to these distributions
that follow the procedure described in Section~\ref{sec:sel_SL}
give the signal yields
 $N_{\rm observed}$  
for each $q^2$ bin.
In these figures, the blue  solid lines show the best fit to the data,
while the red dashed lines show the background.
In these fits, the background normalizations are left free.

To account for detection efficiency and detector resolution,
the number of events $N^i_{\rm observed}$ observed in the $i$th $q^2$ bin is
extracted from the relation
\begin{equation}
N^{i}_{\rm observed}=\sum_{j=1}^{N_{\rm bins}}\epsilon_{ij}N^{j}_{\rm produced},
\label{relation_between_obs_and_prd}
\end{equation}
where $\epsilon_{ij}$ is the overall efficiency matrix that describes the efficiency
and
migration
across $q^{2}$ bins.

The efficiency matrix element $\epsilon_{ij}$ is obtained by
\begin{equation}
 \epsilon_{ij} = \frac{N^{\rm reconstructed}_{ij}}{N^{\rm generated}_j}
\frac{1}{\epsilon_{\rm tag}}
f_{{\rm corr},ij}^{\rm trk+PID},
\end{equation}
where $N^{\rm reconstructed}_{ij}$ is the number of signal Monte Carlo
events generated in the $j$th $q^2$ bin
and reconstructed in the $i$th $q^2$ bin,
$N^{\rm generated}_j$ is the total number of the signal Monte Carlo
events which are generated in the $j$th $q^2$ bin,
$\epsilon_{\rm tag}$ is the single $\bar D^0$ tag efficiency,
and $f_{{\rm corr},ij}^{\rm trk+PID}$ is the efficiency correction matrix
for correcting the Monte Carlo deviations for tracking and particle identification efficiencies
of each element of the efficiency matrix described above.

Table~\ref{tab_eff_matrix_K_all} presents the average overall
efficiency matrix for the $D^0\to K^-e^+\nu_e$ mode.  To produce this
average overall efficiency matrix, we combine the efficiency matrices for
each tag mode weighted by its yields shown in Tab.~\ref{Tab:d_Ntag}.
The diagonal elements of the matrix give the overall efficiency for
$D^0\rightarrow K^-e^+\nu_e$ decays to be reconstructed in the correct
$q^2$ bin in the recoil of the single $\bar D^0$ tags, while the
neighboring off-diagonal elements of the matrix give the overall
efficiency for cross feed between different $q^2$ bins.  Similarly,
Table~\ref{tab_eff_matrix_pi_all} presents the average overall
efficiency matrix for the $D^0\to \pi^-e^+\nu_e$ channel.

\begin{table*}
\caption{Weighted efficiency matrix $\epsilon_{ij}$
(in percent) for $D^{0}\to K^{-}e^{+}\nu_{e}$.
The column gives the true $q^{2}$ bin $j$, while the row gives the
reconstructed $q^{2}$ bin $i$.}
\label{tab_eff_matrix_K_all}
 \begin{ruledtabular}
 \begin{tabular}{c|dddddddddddddddddd}
  $\epsilon_{ij}$ & 1 & 2 & 3 & 4 & 5 & 6 & 7 & 8 & 9 & 10 & 11 & 12 & 13 & 14 & 15 & 16 & 17 & 18 \\
  \hline
  ~1 &  71.62 &   5.39 &   0.70 &   0.31 &   0.11 &   0.03 &   0.01 &   0.00 &   0.01 &   0.00 &   0.00 &   0.00 &   0.00 &   0.00 &   0.00 &   0.00 &   0.00 &   0.00 \\
  ~2 &   2.54 &  65.14 &   6.31 &   0.86 &   0.42 &   0.15 &   0.02 &   0.01 &   0.01 &   0.00 &   0.00 &   0.00 &   0.00 &   0.00 &   0.00 &   0.00 &   0.00 &   0.00 \\
  ~3 &   0.05 &   3.19 &  62.49 &   6.53 &   0.84 &   0.37 &   0.10 &   0.05 &   0.01 &   0.02 &   0.00 &   0.00 &   0.00 &   0.01 &   0.00 &   0.02 &   0.00 &   0.00 \\
  ~4 &   0.02 &   0.09 &   3.61 &  61.47 &   7.07 &   0.79 &   0.31 &   0.09 &   0.01 &   0.00 &   0.01 &   0.01 &   0.00 &   0.00 &   0.00 &   0.00 &   0.00 &   0.00 \\
  ~5 &   0.01 &   0.02 &   0.11 &   3.64 &  60.77 &   6.93 &   0.88 &   0.29 &   0.08 &   0.01 &   0.01 &   0.03 &   0.00 &   0.00 &   0.00 &   0.00 &   0.00 &   0.00 \\
  ~6 &   0.00 &   0.03 &   0.05 &   0.10 &   3.98 &  61.58 &   6.81 &   0.80 &   0.27 &   0.08 &   0.01 &   0.03 &   0.00 &   0.00 &   0.01 &   0.00 &   0.00 &   0.00 \\
  ~7 &   0.01 &   0.01 &   0.01 &   0.05 &   0.14 &   4.16 &  60.75 &   6.71 &   0.66 &   0.27 &   0.04 &   0.02 &   0.02 &   0.00 &   0.00 &   0.00 &   0.00 &   0.00 \\
  ~8 &   0.00 &   0.00 &   0.02 &   0.01 &   0.05 &   0.16 &   4.12 &  60.04 &   6.70 &   0.75 &   0.22 &   0.04 &   0.04 &   0.02 &   0.00 &   0.00 &   0.00 &   0.00 \\
  ~9 &   0.00 &   0.00 &   0.01 &   0.03 &   0.01 &   0.05 &   0.18 &   4.12 &  60.61 &   6.64 &   0.75 &   0.17 &   0.03 &   0.03 &   0.01 &   0.00 &   0.00 &   0.00 \\
  10 &   0.00 &   0.01 &   0.00 &   0.00 &   0.02 &   0.05 &   0.05 &   0.17 &   4.01 &  59.99 &   6.38 &   0.62 &   0.12 &   0.04 &   0.03 &   0.00 &   0.00 &   0.00 \\
  11 &   0.00 &   0.00 &   0.01 &   0.00 &   0.01 &   0.01 &   0.04 &   0.09 &   0.20 &   3.91 &  59.97 &   5.84 &   0.62 &   0.10 &   0.01 &   0.02 &   0.00 &   0.00 \\
  12 &   0.00 &   0.00 &   0.00 &   0.00 &   0.00 &   0.01 &   0.02 &   0.03 &   0.05 &   0.15 &   3.90 &  59.02 &   5.61 &   0.46 &   0.09 &   0.04 &   0.00 &   0.00 \\
  13 &   0.00 &   0.00 &   0.00 &   0.00 &   0.00 &   0.00 &   0.00 &   0.00 &   0.03 &   0.12 &   0.20 &   3.82 &  59.59 &   5.15 &   0.47 &   0.11 &   0.03 &   0.00 \\
  14 &   0.00 &   0.00 &   0.00 &   0.00 &   0.00 &   0.00 &   0.00 &   0.00 &   0.02 &   0.03 &   0.10 &   0.23 &   3.63 &  57.71 &   5.06 &   0.34 &   0.00 &   0.00 \\
  15 &   0.00 &   0.00 &   0.00 &   0.00 &   0.00 &   0.00 &   0.00 &   0.00 &   0.01 &   0.01 &   0.03 &   0.07 &   0.18 &   3.47 &  56.72 &   4.57 &   0.17 &   0.05 \\
  16 &   0.00 &   0.00 &   0.00 &   0.00 &   0.00 &   0.00 &   0.00 &   0.00 &   0.00 &   0.01 &   0.00 &   0.02 &   0.03 &   0.28 &   2.82 &  54.82 &   3.66 &   0.17 \\
  17 &   0.00 &   0.00 &   0.00 &   0.00 &   0.00 &   0.00 &   0.00 &   0.01 &   0.00 &   0.00 &   0.01 &   0.00 &   0.02 &   0.04 &   0.21 &   2.66 &  49.34 &   2.10 \\
  18 &   0.00 &   0.01 &   0.00 &   0.00 &   0.00 &   0.00 &   0.00 &   0.00 &   0.00 &   0.00 &   0.00 &   0.00 &   0.01 &   0.01 &   0.02 &   0.05 &   1.95 &  38.30 \\
 \end{tabular}
 \end{ruledtabular}
\end{table*}

\begin{table*}
\caption{Weighted efficiency matrix $\epsilon_{ij}$
(in percent) for $D^{0}\to \pi^{-}e^{+}\nu_{e}$.
The column gives the true $q^{2}$ bin $j$, while the row gives the reconstructed $q^{2}$ bin $i$.}
\label{tab_eff_matrix_pi_all}
 \begin{ruledtabular}
 \begin{tabular}{c|dddddddddddddd}
  $\epsilon_{ij}$ & 1 & 2 & 3 & 4 & 5 & 6 & 7 & 8 & 9 & 10 & 11 & 12 & 13 & 14  \\
  \hline
  ~1 &  71.86 &   4.55 &   0.57 &   0.04 &   0.01 &   0.00 &   0.00 &   0.00 &   0.00 &   0.00 &   0.00 &   0.00 &   0.00 &   0.00 \\
  ~2 &   1.71 &  67.50 &   5.18 &   0.47 &   0.03 &   0.01 &   0.00 &   0.00 &   0.00 &   0.00 &   0.01 &   0.00 &   0.00 &   0.03 \\
  ~3 &   0.05 &   2.11 &  67.78 &   5.22 &   0.32 &   0.03 &   0.00 &   0.01 &   0.00 &   0.00 &   0.00 &   0.00 &   0.00 &   0.00 \\
  ~4 &   0.02 &   0.07 &   2.42 &  69.05 &   5.30 &   0.25 &   0.01 &   0.00 &   0.00 &   0.00 &   0.00 &   0.00 &   0.01 &   0.00 \\
  ~5 &   0.00 &   0.03 &   0.10 &   2.64 &  69.00 &   5.11 &   0.26 &   0.02 &   0.01 &   0.00 &   0.00 &   0.00 &   0.00 &   0.00 \\
  ~6 &   0.01 &   0.02 &   0.03 &   0.14 &   2.57 &  70.20 &   4.85 &   0.17 &   0.02 &   0.00 &   0.00 &   0.00 &   0.00 &   0.00 \\
  ~7 &   0.01 &   0.02 &   0.04 &   0.05 &   0.15 &   2.67 &  71.01 &   4.46 &   0.13 &   0.01 &   0.00 &   0.00 &   0.00 &   0.00 \\
  ~8 &   0.01 &   0.02 &   0.02 &   0.04 &   0.10 &   0.23 &   2.75 &  71.32 &   4.38 &   0.11 &   0.00 &   0.00 &   0.00 &   0.00 \\
  ~9 &   0.01 &   0.01 &   0.01 &   0.03 &   0.04 &   0.12 &   0.26 &   2.70 &  70.75 &   3.73 &   0.08 &   0.01 &   0.00 &   0.00 \\
  10 &   0.00 &   0.00 &   0.00 &   0.00 &   0.03 &   0.06 &   0.16 &   0.29 &   2.55 &  69.48 &   3.50 &   0.14 &   0.02 &   0.00 \\
  11 &   0.00 &   0.00 &   0.00 &   0.00 &   0.00 &   0.02 &   0.07 &   0.12 &   0.32 &   2.72 &  69.08 &   3.28 &   0.03 &   0.02 \\
  12 &   0.00 &   0.00 &   0.00 &   0.00 &   0.00 &   0.00 &   0.01 &   0.05 &   0.19 &   0.38 &   2.63 &  65.19 &   3.09 &   0.01 \\
  13 &   0.00 &   0.00 &   0.00 &   0.00 &   0.01 &   0.00 &   0.00 &   0.01 &   0.01 &   0.13 &   0.41 &   2.49 &  64.07 &   2.81 \\
  14 &   0.00 &   0.00 &   0.00 &   0.00 &   0.00 &   0.00 &   0.01 &   0.00 &   0.01 &   0.02 &   0.08 &   0.33 &   2.29 &  66.78 \\
 \end{tabular}
 \end{ruledtabular}
\end{table*}

The number of $D^0\rightarrow K^-(\pi^-)e^+\nu_e$ semileptonic decay events
produced with $q^2$ filled in the $i$th $q^{2}$ bin is obtained from
\begin{equation}
\label{eq04}
 N_{\rm produced}^{i}=\sum_{j}^{N_{\rm bins}}(\epsilon^{-1})_{ij}N_{\rm observed}^{j},
\end{equation}
with a statistical error given by
\begin{equation}
    \sigma_{\rm stat}(N_{\rm produced}^{i})=
       \sqrt{ \sum_{j}^{N_{\rm bins}}\left(\epsilon^{-1}\right)^{2}_{ij}
        \left(\sigma_{\rm stat}(N_{\rm observed}^{j})\right)^{2}},
\label{eq19}
\end{equation}
in which $\sigma_{\rm stat}(N_{\rm observed}^{j})$ is the statistical error of $N_{\rm observed}^{j}$.
The partial width for the $i$th bin is given by
\begin{equation}
\Delta\Gamma_{i}=\frac{N_{\rm produced}^{i}}{\tau_{D^0} N_{\rm tag}},
\label{eq20}
\end{equation}
where $\tau_{D^0}$ is the lifetime of the $D^0$ meson and $N_{\rm tag}$ is the number of the single
$\bar D^0$ tags.

The numbers of the signal events and $q^2$-dependent partial widths
for $D^0 \rightarrow K^-e^+\nu_e$ and $D^0 \rightarrow \pi^-e^+\nu_e$
are summarized in Table~\ref{tab_Nobs_q2_K} and Table~\ref{tab_Nobs_q2_pi}, respectively, where the errors are statistical only.

\subsection{Fitting partial decay rates to extract form factors}
\label{lab_sec_6_2}

To extract the form-factor parameters,
we fit
the theoretical predictions of the rates to the measured partial decay rates.
Taking into account the correlations of the measured partial decay
rates among $q^2$ bins, the $\chi^2$
to be minimized
is defined as
\begin{eqnarray}\label{eq_chi2_fit}
  \chi^2 &=& \sum_{i,j=1}^{N_{\rm bins}}
	  (\Delta\Gamma_i^{\rm measured}-\Delta\Gamma_i^{\rm expected})
	  C^{-1}_{ij}
(\Delta\Gamma_j^{\rm measured}
\nonumber\\
& & -\Delta\Gamma_j^{\rm expected})),
\end{eqnarray}
where $\Delta\Gamma_i^{\rm measured}$ is the measured partial decay
rate in $i$th $q^2$ bin,
$ C_{ij}^{-1}$ is the inverse of the covariance matrix
$ C_{ij}$ which accounts for the correlations between the
measured partial decay rates in different $q^2$ bins,
and $N_{\rm bins}$ is the number of $q^2$ bins.
The expected partial decay rate in the $i$th $q^2$ bin is given by
\begin{equation}\label{partail_width_vs_form_factor}
  \Delta\Gamma_i^{\rm expected}
  = \int_{q^{2}_{{\rm min}(i)}}^{q^{2}_{{\rm max}(i)}}
    \frac{G^2_F|V_{cs(d)}|^2}{24\pi^3}|\vec p_{K(\pi)}|^3|f_+^{K(\pi)}(q^2)|^2 dq^2,
\end{equation}
where ${q^2_{{\rm min}(i)}}$ and ${q^2_{{\rm max}(i)}}$ are the lower and higher boundaries
of the $q^2$ bin $i$, respectively.
In the fits, all parameters of the form-factor parameterizations are left free.

We separate the covariance matrix into two parts, one is the statistical covariance matrix $ C_{ij}^{\rm stat}$
and the other is the systematic covariance matrix $ C_{ij}^{\rm sys}$.
The statistical covariance matrix is determined by
\begin{equation}
 C^{\rm stat}_{ij} = \left(\frac{1}{\tau_{D^0} N_{\rm tag}}\right)^{2} \sum_{\alpha} \epsilon^{-1}_{i \alpha}
                     \epsilon^{-1}_{j \alpha} \left(\sigma(N_{\rm observed}^{\alpha})\right)^2.
\end{equation}
Table~\ref{tab_cov_stat_Kenu} and Table~\ref{tab_cov_stat_pienu} give
the statistical correlation matrix and relative statistical
uncertainties of the measured partial decay rates
for $D^0 \rightarrow K^-e^+\nu_e$ and $D^0 \rightarrow \pi^-e^+\nu_e$ decays, respectively.

Inserting the inverse statistical covariance matrix $( C^{\rm stat})^{-1}$
into Eq.~(\ref{eq_chi2_fit}),
replacing the form factor $f^{K(\pi)}_+(q^2)$ in Eq.~(\ref{partail_width_vs_form_factor})
with different form-factor parameterizations discussed in the Section~\ref{sec:th},
and fitting to the measured partial decay rates
yields the product of $f_+^{K(\pi)}(0)$ and $|V_{cs(d)}|$
as well as the parameters of the form factor.

\begin{table*}
\caption{Statistical correlation matrix and
relative statistical uncertainty of the measured partial decay rate in each $q^2$ bin
for $D^0\to K^-e^+\nu_e$.}
\label{tab_cov_stat_Kenu}
 \begin{ruledtabular}
 \begin{tabular}{ldddddddddd}
  $q^2$ bin & 1 & 2 & 3 & 4 & 5 & 6 & 7 & 8 & 9 & 10 \\
  correlation $\rho_{ij}$
&   1.000  &          &          &          &          &          &          &          &          &          \\
&  -0.115  &   1.000  &          &          &          &          &          &          &          &          \\
&   0.003  &  -0.146  &   1.000  &          &          &          &          &          &          &          \\
&  -0.003  &   0.005  &  -0.159  &   1.000  &          &          &          &          &          &          \\
&  -0.001  &  -0.005  &   0.007  &  -0.171  &   1.000  &          &          &          &          &          \\
&   0.000  &  -0.001  &  -0.005  &   0.011  &  -0.172  &   1.000  &          &          &          &          \\
&   0.000  &   0.000  &   0.000  &  -0.004  &   0.009  &  -0.175  &   1.000  &          &          &          \\
&   0.000  &   0.000  &  -0.001  &   0.000  &  -0.004  &   0.010  &  -0.174  &   1.000  &          &          \\
&   0.000  &   0.000  &   0.000  &   0.000  &   0.000  &  -0.004  &   0.011  &  -0.174  &   1.000  &          \\
&   0.000  &   0.000  &   0.000  &   0.000  &   0.000  &  -0.001  &  -0.004  &   0.010  &  -0.171  &   1.000  \\
&   0.000  &   0.000  &   0.000  &   0.000  &   0.000  &   0.000  &   0.000  &  -0.003  &   0.008  &  -0.166  \\
&   0.000  &   0.000  &   0.000  &   0.000  &   0.000  &   0.000  &   0.000  &   0.000  &  -0.002  &   0.010  \\
&   0.000  &   0.000  &   0.000  &   0.000  &   0.000  &   0.000  &   0.000  &   0.000  &   0.000  &  -0.003  \\
&   0.000  &   0.000  &   0.000  &   0.000  &   0.000  &   0.000  &   0.000  &   0.000  &  -0.001  &   0.000  \\
&   0.000  &   0.000  &   0.000  &   0.000  &   0.000  &   0.000  &   0.000  &   0.000  &   0.000  &   0.000  \\
&   0.000  &   0.000  &   0.000  &   0.000  &   0.000  &   0.000  &   0.000  &   0.000  &   0.000  &   0.000  \\
&   0.000  &   0.000  &   0.000  &   0.000  &   0.000  &   0.000  &   0.000  &   0.000  &   0.000  &   0.000  \\
&   0.000  &  -0.001  &   0.000  &   0.000  &   0.000  &   0.000  &   0.000  &   0.000  &   0.000  &   0.000  \\
stat. uncert. &  1.31\%  &  1.41\%  &  1.49\%  &  1.60\%  &  1.61\%  &  1.74\%  &  1.77\%  &  1.90\%  &  1.97\%  &  2.08\%  \\
\hline
$q^2$ bin & 11 & 12 & 13 & 14 & 15 & 16 & 17 & 18 \\
correlation  $\rho_{ij}$
&   1.000  &          &          &          &          &          &          &          \\
&  -0.159  &   1.000  &          &          &          &          &          &          \\
&   0.007  &  -0.154  &   1.000  &          &          &          &          &          \\
&  -0.002  &   0.007  &  -0.146  &   1.000  &          &          &          &          \\
&   0.000  &  -0.002  &   0.006  &  -0.145  &   1.000  &          &          &          \\
&   0.000  &  -0.001  &  -0.001  &   0.003  &  -0.127  &   1.000  &          &          \\
&   0.000  &   0.000  &  -0.001  &   0.000  &   0.004  &  -0.119  &   1.000  &          \\
&   0.000  &   0.000  &   0.000  &   0.000  &  -0.001  &   0.006  &  -0.098  &   1.000  \\
stat. uncert.
&  2.20\%  &  2.42\%  &  2.60\%  &  2.95\%  &  3.39\%  &  3.92\%  &  5.11\%  &  9.12\%   \\
 \end{tabular}
 \end{ruledtabular}
\end{table*}

\begin{table*}
\caption{Statistical correlation matrix and
relative statistical uncertainty of the measured partial decay rate in each $q^2$ bin
for $D^0\to \pi^-e^+\nu_e$.}
\label{tab_cov_stat_pienu}
 \begin{ruledtabular}
 \begin{tabular}{ldddddddddd}
  $q^2$ bin & 1 & 2 & 3 & 4 & 5 & 6 & 7 & 8 & 9 & 10 \\
  correlation  $\rho_{ij}$
&  1.000 &        &        &        &        &        &        &        &        &        \\
& -0.087 &  1.000 &        &        &        &        &        &        &        &        \\
& -0.001 & -0.105 &  1.000 &        &        &        &        &        &        &        \\
&  0.000 &  0.002 & -0.110 &  1.000 &        &        &        &        &        &        \\
&  0.000 &  0.000 &  0.004 & -0.114 &  1.000 &        &        &        &        &        \\
&  0.000 &  0.000 & -0.001 &  0.004 & -0.108 &  1.000 &        &        &        &        \\
&  0.000 &  0.000 &  0.000 & -0.001 &  0.003 & -0.104 &  1.000 &        &        &        \\
&  0.000 &  0.000 &  0.000 &  0.000 & -0.001 &  0.003 & -0.099 &  1.000 &        &        \\
&  0.000 &  0.000 &  0.000 &  0.000 &  0.000 & -0.002 &  0.002 & -0.098 &  1.000 &        \\
&  0.000 &  0.000 &  0.000 &  0.000 &  0.000 & -0.001 & -0.002 &  0.001 & -0.087 &  1.000 \\
&  0.000 &  0.000 &  0.000 &  0.000 &  0.000 &  0.000 & -0.001 & -0.002 &  0.000 & -0.088 \\
&  0.000 &  0.000 &  0.000 &  0.000 &  0.000 &  0.000 &  0.000 & -0.001 & -0.004 & -0.003 \\
&  0.000 &  0.000 &  0.000 &  0.000 &  0.000 &  0.000 &  0.000 &  0.000 &  0.000 & -0.003 \\
&  0.000 &  0.000 &  0.000 &  0.000 &  0.000 &  0.000 &  0.000 &  0.000 &  0.000 &  0.000 \\
stat. uncert.
& 4.05\% & 4.58\% & 4.93\% & 4.77\% & 4.75\% & 5.07\% & 5.39\% & 5.94\% & 5.93\% & 6.80\% \\
\hline
$q^2$ bin & 11 & 12 & 13 & 14  \\
correlation  $\rho_{ij}$
&  1.000 &        &        &        \\
& -0.087 &  1.000 &        &        \\
& -0.003 & -0.086 &  1.000 &        \\
& -0.001 & -0.002 & -0.077 &  1.000 \\
stat. uncert.
& 7.64\% & 9.45\% & 12.96\% & 13.95\% \\
 \end{tabular}
 \end{ruledtabular}
\end{table*}

\subsection{Systematic uncertainties in form factor measurements}
\label{lab_sec_6_3}

\subsubsection{Systematic covariance matrix}

For each source of systematic uncertainty, an $N_{\rm bins} \times N_{\rm bins}$ covariance matrix
is estimated.
The total systematic covariance matrix is obtained by summing all these matrices.

\begin{enumerate}[(1)]
  \item {\bf\boldmath\em Number of $\bar D^0$ tags} \\
  The uncertainties associated with the number of the single $\bar
  D^0$ tags are fully correlated across all $q^2$ bins.
  The systematic covariance contributed from the uncertainty in the number of single $\bar D^0$ tags is calculated by
  \begin{equation}
   C^{\rm sys}_{ij}(N_{\rm tag}) = \Delta \Gamma_{i} \Delta \Gamma_{j} \left(\frac{\sigma(N_{\rm tag})}{N_{\rm tag}}\right)^{2},
  \end{equation}
  where $\sigma(N_{\rm tag})/N_{\rm tag}$ is the relative uncertainty of the number of the single $\bar D^0$ tags.

  \item {\bf\boldmath\em $D^0$ lifetime} \\
  The uncertainty associated with the lifetime of the $D^0$ meson are
  fully correlated across all $q^2$ bins,
  so the systematic covariance is calculated by
  \begin{equation}
   C^{\rm sys}_{ij}(\tau_{D^0}) = \Delta \Gamma_{i} \Delta \Gamma_{j} \left(\frac{\sigma(\tau_{D^0})}{\tau_{D^0}}\right)^{2},
  \end{equation}
  where $\sigma(\tau_{D^0})$ is the uncertainty of the $D^0$ lifetime
  taken from PDG~\cite{pdg2014}.

  \item {\bf\em Monte Carlo statistics} \\
  The systematic uncertainties and correlations in $q^2$ bins due to
  the limited size of the Monte Carlo samples
  used to determine the efficiency matrices are calculated by
  \begin{eqnarray}
    C^{\rm sys}_{ij}({\rm MC~stat.}) &=& \left(\frac{1}{\tau_{D^0} N_{\rm tag}}\right)^{2} \sum_{\alpha \beta}
    (N_{\rm observed}^{\alpha}
    \nonumber\\
    & \times & N_{\rm observed}^{\beta} \mathrm{Cov}(\epsilon^{-1}_{i \alpha},\epsilon^{-1}_{j \beta}) ),
  \end{eqnarray}
where the covariance of the inverse efficiency matrix elements are given
by~\cite{cov_inverse_matrix}
   \begin{equation}
    \mathrm{Cov}(\epsilon^{-1}_{\alpha\beta},\epsilon^{-1}_{ab})
    = \sum_{ij}(\epsilon^{-1}_{\alpha i}\epsilon^{-1}_{ai})
    [\sigma^2(\epsilon_{ij})]^2
       (\epsilon^{-1}_{j \beta}\epsilon^{-1}_{jb}).
   \end{equation}

  \item {\bf\em Form factor structure} \\
In order to estimate the systematic uncertainty associated with
the form factor used to generate signal events in the Monte Carlo simulation,
we re-weight the signal Monte Carlo events so that the $q^2$ spectra
agree with the measured spectra.
We then re-calculate the
partial decay rates with the new efficiency matrices
which are determined using the weighted Monte Carlo events.
The covariance matrix due to this source is assigned via
\begin{equation}
 C^{\rm sys}_{ij}({\rm F.F.})=\delta(\Delta\Gamma_i)\delta(\Delta\Gamma_j),
\end{equation}
where $\delta(\Delta\Gamma_i)$ denotes the change in the measured
partial rate in the $i$th $q^2$ bin.

  \item {\bf\boldmath\em $E_{\gamma,\rm max}$ cut} \\
  We assign systematic uncertainties of $0.10\%$ due to the $E_{\gamma,\rm max}$
  requirement on the selected events in each $q^2$ bin,
  and assume that they are fully correlated between $q^2$ bins.
  The systematic covariance due to this requirement can be obtained by
  \begin{equation}
   C^{\rm sys}_{ij}(E_{\gamma,\rm max})=\sigma(\Delta\Gamma_i)\sigma(\Delta\Gamma_j),
   \end{equation}
  where $\sigma(\Delta\Gamma_i)=0.10\% \times \Delta\Gamma_i$.

  \item {\bf\boldmath\em $U_{\rm miss}$ fits} \\
  The technique of fitting $U_{\rm miss}$ distributions affects the numbers of signal events observed in $q^2$ bins.
  The covariance matrix due to the $U_{\rm miss}$ fits is determined by
  \begin{eqnarray}
   C^{\rm sys}_{ij}(U_{\rm miss} {\rm~ Fit}) &=&\left(\frac{1}{\tau_D N_{\rm tag}}\right)^{2}
   \nonumber\\
   &\times& \sum_{\alpha} \epsilon^{-1}_{i \alpha}
   \epsilon^{-1}_{j \alpha} \left(\sigma_{\alpha}^{\rm Fit}\right)^2,
  \end{eqnarray}
  where $\sigma_{\alpha}^{\rm Fit}$ is the systematic uncertainty of the number of the signal events observed in the bin $\alpha$
  due to fitting $U_{\rm miss}$ distribution,
  evaluated as described in Sect.~\ref{sec:syst_umiss}.

  \item {\bf\em Tracking and PID efficiencies} \\
  The covariance matrices for the systematic uncertainties associated with the tracking efficiencies and
  the particle identification efficiencies for the charged particles are obtained in the following way.
  We first vary the correction coefficients for tracking (PID) efficiencies by $\pm 1\sigma$, then remeasure
  the partial decay rates using the efficiency matrices obtained from the re-corrected signal Monte Carlo events.
  The covariance matrix due to this source is assigned via
  \begin{equation}
   C^{\rm sys}_{ij}(\rm Tracking, PID)=\delta(\Delta\Gamma_i)\delta(\Delta\Gamma_j),
  \end{equation}
  where $\delta(\Delta\Gamma_i)$ denotes the change in the measured partial decay rate in the $i$th $q^2$ bin.

  \item {\bf\em FSR recovery} \\
  To estimate the systematic covariance matrix associated with the FSR recovery of the positron momentum,
  we remeasure the partial decay rates without the FSR recovery.
  The covariance matrix due to this source is assigned via
  \begin{equation}
   C^{\rm sys}_{ij}(\rm FSR)=\delta(\Delta\Gamma_i)\delta(\Delta\Gamma_j),
   \end{equation}
  where $\delta(\Delta\Gamma_i)$ denotes the change in the measured partial decay rate in $i$th $q^2$ bin.

  \item {\bf\em Single tag cancelation} \\
  We take the systematic uncertainties associated with single tag cancelation
  as 0.12\% in each $q^2$ bin, and assume they are fully correlated between
  different $q^2$ bins.
\end{enumerate}

The total systematic correlation matrix and  relative systematic uncertainties
for measurements of the partial decay rates of the two semileptonic decays of
$D^0\to K^-e^+\nu_e$ and $D^0\to \pi^-e^+\nu_e$
are presented in the Table~\ref{tab_cov_sys_Kenu} and Table~\ref{tab_cov_sys_pienu}, respectively.

\begin{table*}
\caption{Systematic correlation matrix and
relative systematic uncertainty of the measured partial decay rate in each $q^2$ bin
for $D^0\to K^-e^+\nu_e$.}
\label{tab_cov_sys_Kenu}
 \begin{ruledtabular}
 \begin{tabular}{ldddddddddd}
  $q^2$ bin & 1 & 2 & 3 & 4 & 5 & 6 & 7 & 8 & 9 & 10 \\
  correlation $\rho_{ij}$
&  1.000 &        &        &        &        &        &        &        &        &        \\
&  0.284 &  1.000 &        &        &        &        &        &        &        &        \\
&  0.356 &  0.407 &  1.000 &        &        &        &        &        &        &        \\
&  0.350 &  0.482 &  0.406 &  1.000 &        &        &        &        &        &        \\
&  0.354 &  0.481 &  0.496 &  0.412 &  1.000 &        &        &        &        &        \\
&  0.350 &  0.477 &  0.485 &  0.499 &  0.419 &  1.000 &        &        &        &        \\
&  0.353 &  0.481 &  0.492 &  0.496 &  0.513 &  0.419 &  1.000 &        &        &        \\
&  0.350 &  0.477 &  0.488 &  0.496 &  0.506 &  0.509 &  0.426 &  1.000 &        &        \\
&  0.340 &  0.465 &  0.477 &  0.485 &  0.498 &  0.493 &  0.506 &  0.414 &  1.000 &        \\
&  0.332 &  0.454 &  0.464 &  0.472 &  0.482 &  0.479 &  0.483 &  0.489 &  0.388 &  1.000 \\
&  0.317 &  0.433 &  0.442 &  0.449 &  0.457 &  0.455 &  0.461 &  0.458 &  0.456 &  0.345 \\
&  0.297 &  0.406 &  0.414 &  0.419 &  0.428 &  0.425 &  0.429 &  0.428 &  0.419 &  0.413 \\
&  0.282 &  0.386 &  0.395 &  0.400 &  0.409 &  0.407 &  0.411 &  0.410 &  0.404 &  0.390 \\
&  0.264 &  0.361 &  0.370 &  0.377 &  0.386 &  0.384 &  0.388 &  0.388 &  0.381 &  0.369 \\
&  0.222 &  0.304 &  0.312 &  0.317 &  0.325 &  0.323 &  0.326 &  0.326 &  0.321 &  0.310 \\
&  0.211 &  0.288 &  0.297 &  0.303 &  0.312 &  0.310 &  0.314 &  0.315 &  0.310 &  0.299 \\
&  0.170 &  0.233 &  0.241 &  0.248 &  0.256 &  0.255 &  0.259 &  0.259 &  0.257 &  0.247 \\
&  0.120 &  0.163 &  0.172 &  0.180 &  0.188 &  0.187 &  0.191 &  0.192 &  0.192 &  0.183 \\
syst. uncert.
& 1.67\% & 1.21\% & 1.20\% & 1.24\% & 1.23\% & 1.24\% & 1.25\% & 1.25\% & 1.30\% & 1.32\% \\
\hline
$q^2$ bin & 11 & 12 & 13 & 14 & 15 & 16 & 17 & 18 \\
correlation $\rho_{ij}$
&  1.000 &        &        &        &        &        &        &        \\
&  0.291 &  1.000 &        &        &        &        &        &        \\
&  0.377 &  0.246 &  1.000 &        &        &        &        &        \\
&  0.349 &  0.331 &  0.214 &  1.000 &        &        &        &        \\
&  0.295 &  0.273 &  0.268 &  0.140 &  1.000 &        &        &        \\
&  0.283 &  0.263 &  0.253 &  0.243 &  0.101 &  1.000 &        &        \\
&  0.233 &  0.216 &  0.208 &  0.197 &  0.170 &  0.063 &  1.000 &        \\
&  0.171 &  0.157 &  0.152 &  0.146 &  0.122 &  0.127 &  0.019 &  1.000 \\
syst. uncert.
& 1.36\% & 1.42\% & 1.50\% & 1.62\% & 1.93\% & 2.04\% & 2.78\% & 4.66\% \\
 \end{tabular}
 \end{ruledtabular}
\end{table*}

\begin{table*}
\caption{Systematic correlation matrix and
relative systematic uncertainty of the measured partial decay rate in each $q^2$ bin
for $D^0\to \pi^-e^+\nu_e$.}
\label{tab_cov_sys_pienu}
 \begin{ruledtabular}
 \begin{tabular}{ldddddddddd}
  $q^2$ bin  & 1 & 2 & 3 & 4 & 5 & 6 & 7 & 8 & 9 & 10 \\
correlation $\rho_{ij}$
&  1.000 &        &        &        &        &        &        &        &        &        \\
&  0.271 &  1.000 &        &        &        &        &        &        &        &        \\
&  0.286 &  0.200 &  1.000 &        &        &        &        &        &        &        \\
&  0.305 &  0.299 &  0.183 &  1.000 &        &        &        &        &        &        \\
&  0.252 &  0.246 &  0.220 &  0.137 &  1.000 &        &        &        &        &        \\
&  0.276 &  0.272 &  0.240 &  0.263 &  0.134 &  1.000 &        &        &        &        \\
&  0.200 &  0.197 &  0.174 &  0.187 &  0.158 &  0.077 &  1.000 &        &        &        \\
&  0.206 &  0.207 &  0.183 &  0.197 &  0.163 &  0.187 &  0.048 &  1.000 &        &        \\
&  0.174 &  0.176 &  0.156 &  0.168 &  0.140 &  0.155 &  0.115 &  0.031 &  1.000 &        \\
&  0.201 &  0.205 &  0.181 &  0.195 &  0.162 &  0.182 &  0.129 &  0.143 &  0.043 &  1.000 \\
&  0.169 &  0.173 &  0.153 &  0.165 &  0.138 &  0.154 &  0.111 &  0.119 &  0.101 &  0.043 \\
&  0.132 &  0.135 &  0.120 &  0.129 &  0.108 &  0.121 &  0.088 &  0.094 &  0.077 &  0.093 \\
&  0.144 &  0.147 &  0.130 &  0.141 &  0.117 &  0.133 &  0.096 &  0.103 &  0.089 &  0.101 \\
&  0.110 &  0.107 &  0.093 &  0.102 &  0.086 &  0.097 &  0.069 &  0.074 &  0.063 &  0.074 \\
syst. uncert.
& 1.58\% & 1.47\% & 1.65\% & 1.53\% & 1.85\% & 1.66\% & 2.28\% & 2.09\% & 2.42\% & 2.10\% \\
\hline
$q^2$ bin & 11 & 12 & 13 & 14  \\
correlation $\rho_{ij}$
&  1.000 &        &        &        \\
&  0.002 &  1.000 &        &        \\
&  0.083 & -0.020 &  1.000 &        \\
&  0.062 &  0.047 & -0.023 &  1.000 \\
syst. uncert.
& 2.43\%  & 3.09\%  & 2.98\%  & 4.92\%    \\
 \end{tabular}
 \end{ruledtabular}
\end{table*}

\subsubsection{Systematic uncertainty in measurements of form factor parameters}

To obtain the systematic uncertainty of the parameters of the form factors obtained from the fits,
we add the matrix elements of the statistical covariance matrix
and systematic covariance matrix together. We then repeat the fits to the partial decay rates.

The central values of the form factor parameters are taken from the results obtained by fitting the data
with the combined statistical and systematic covariance matrix together.
The quadrature difference between the uncertainties of the fit parameters
obtained from the fits with the combined covariance matrix
and the uncertainties of the fit parameters
obtained from the fits with only the statistical covariance matrix
is taken as the systematic error of the measured form factor parameter.

\subsection{Results of form-factor measurements}
\label{lab_sec_6_4}

\begin{figure*}[h!bt]
\begin{minipage}[t]{0.49\linewidth}
\centering
\includegraphics[width=\textwidth]{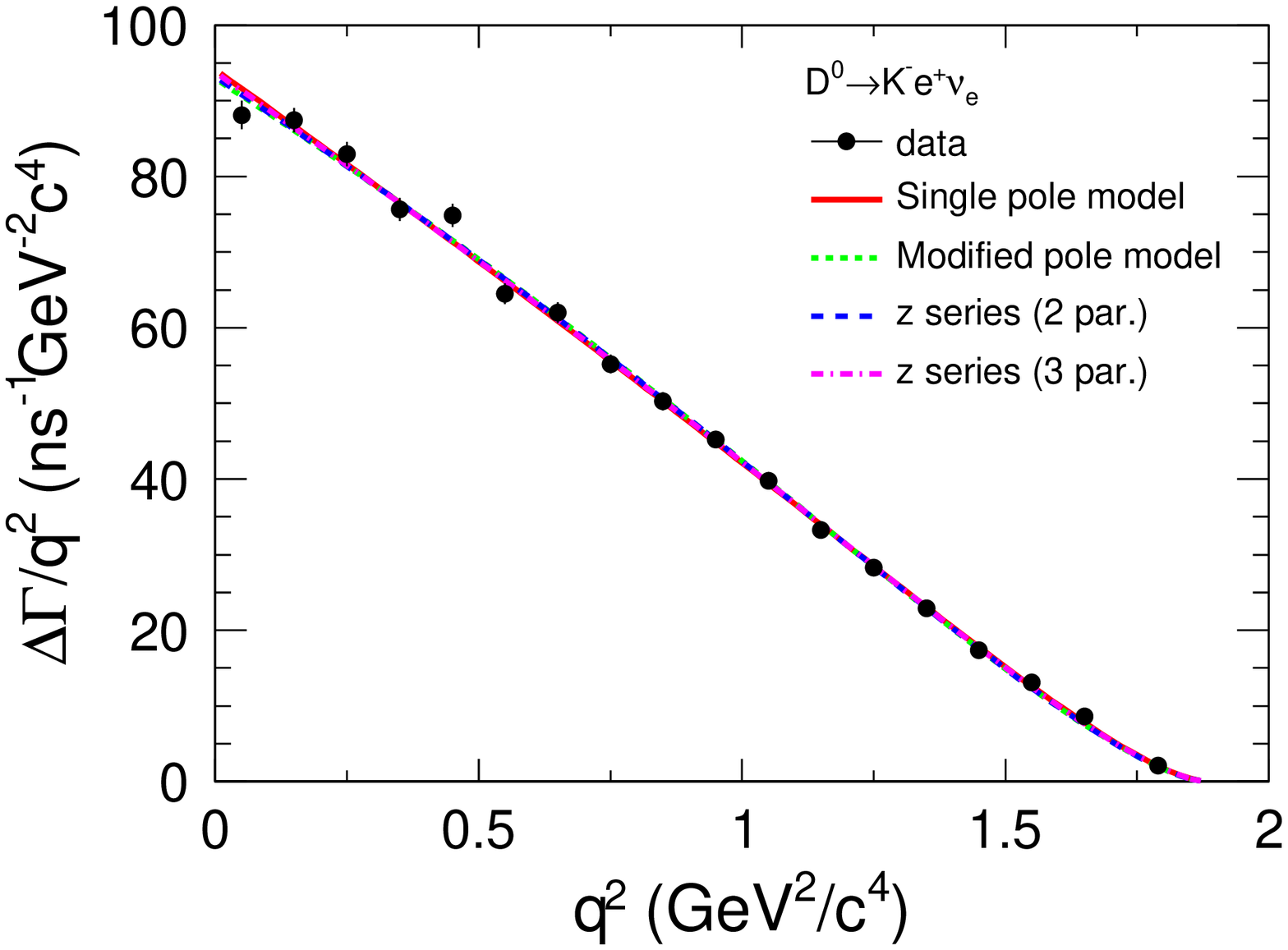}
\caption{
Differential decay rates for $D^0 \rightarrow K^-e^+\nu_e$ as function
of $q^2$.
The dots with error bars show the data
and the lines give the best fits to the data
with different form-factor parameterizations.
}
\label{Form_Factor_Fits_Plots_for_Kev}
\end{minipage}
\hfill
\begin{minipage}[t]{0.49\linewidth}
\centering
\includegraphics[width=\textwidth]{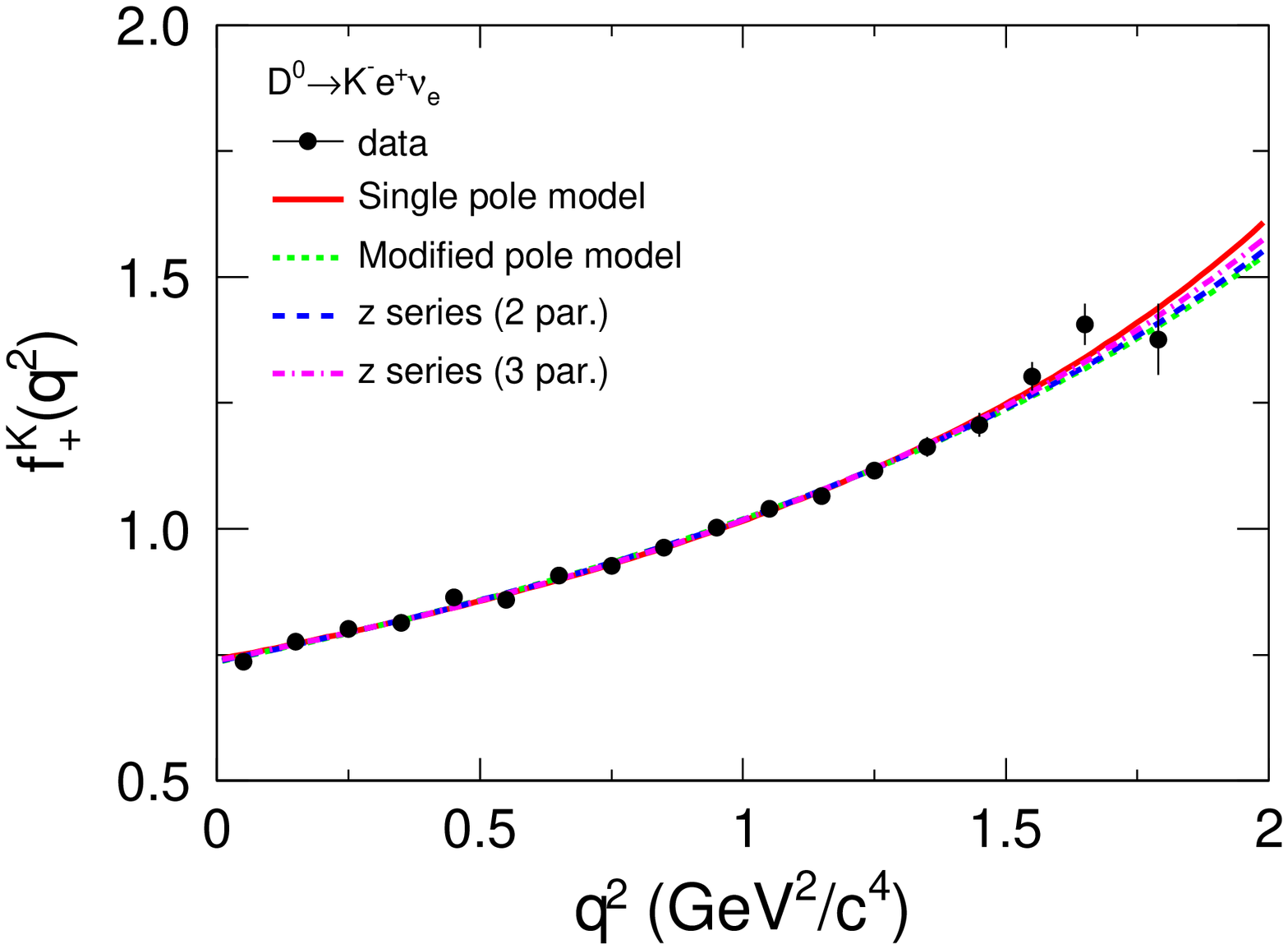}
\caption{Projections on $f_+^K(q^2)$  for $D^{0}\to K^{-}e^{+}\nu_{e}$.}
\label{projections_of_Fits_onto_ff_q2_stat_Kev}
\end{minipage}
\begin{minipage}[t]{0.49\linewidth}
\centering
\includegraphics[width=\textwidth]{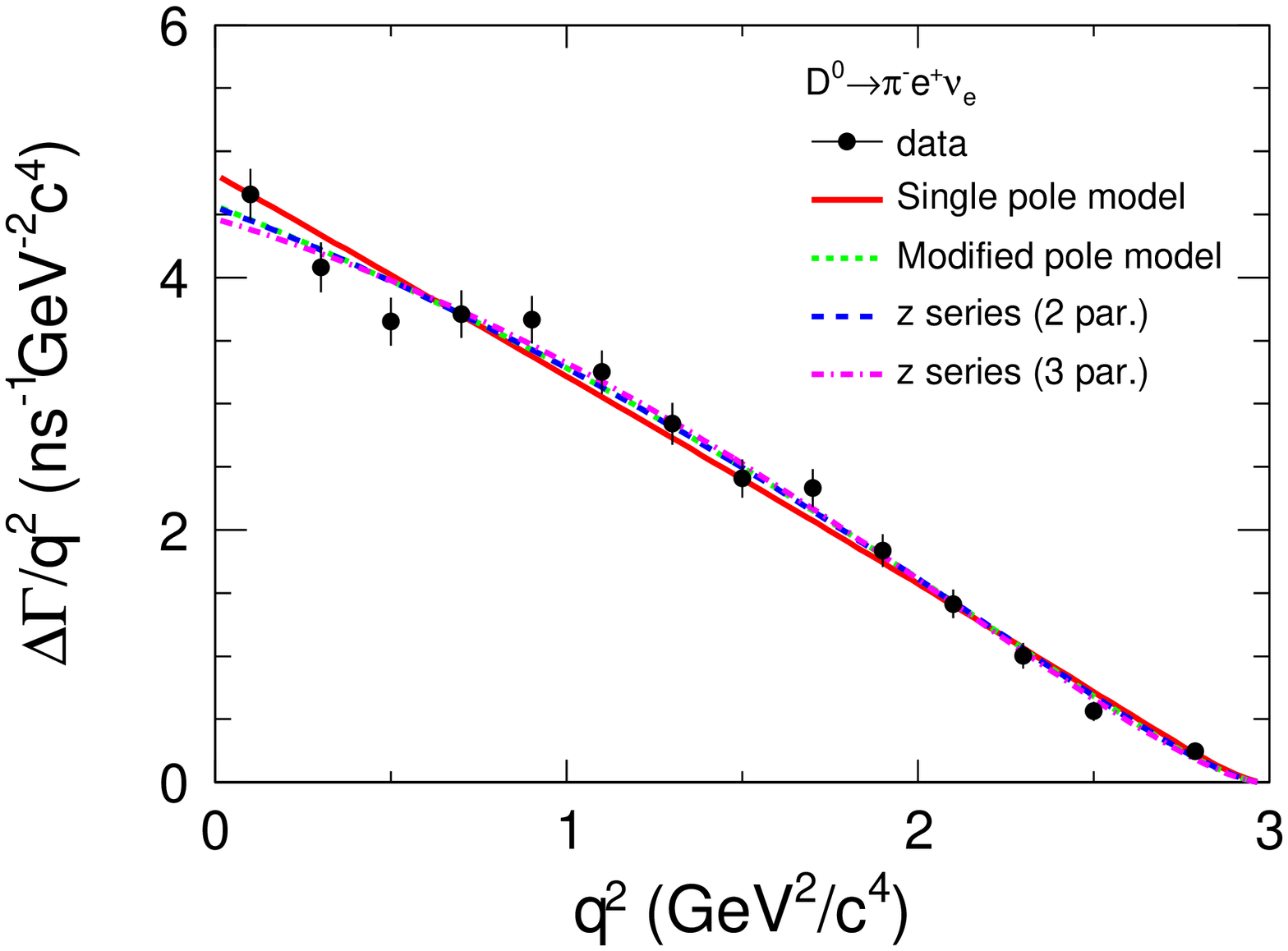}
\caption{
Differential decay rates for $D^0 \rightarrow \pi^-e^+\nu_e$ as
function of $q^2$.
The dots with error bars show the data
and the lines give the best fits to the data
with different form-factor parameterizations.
}
\label{Form_Factor_Fits_Plots_for_piev}
\end{minipage}
\hfill
\begin{minipage}[t]{0.49\linewidth}
\centering
\includegraphics[width=\textwidth]{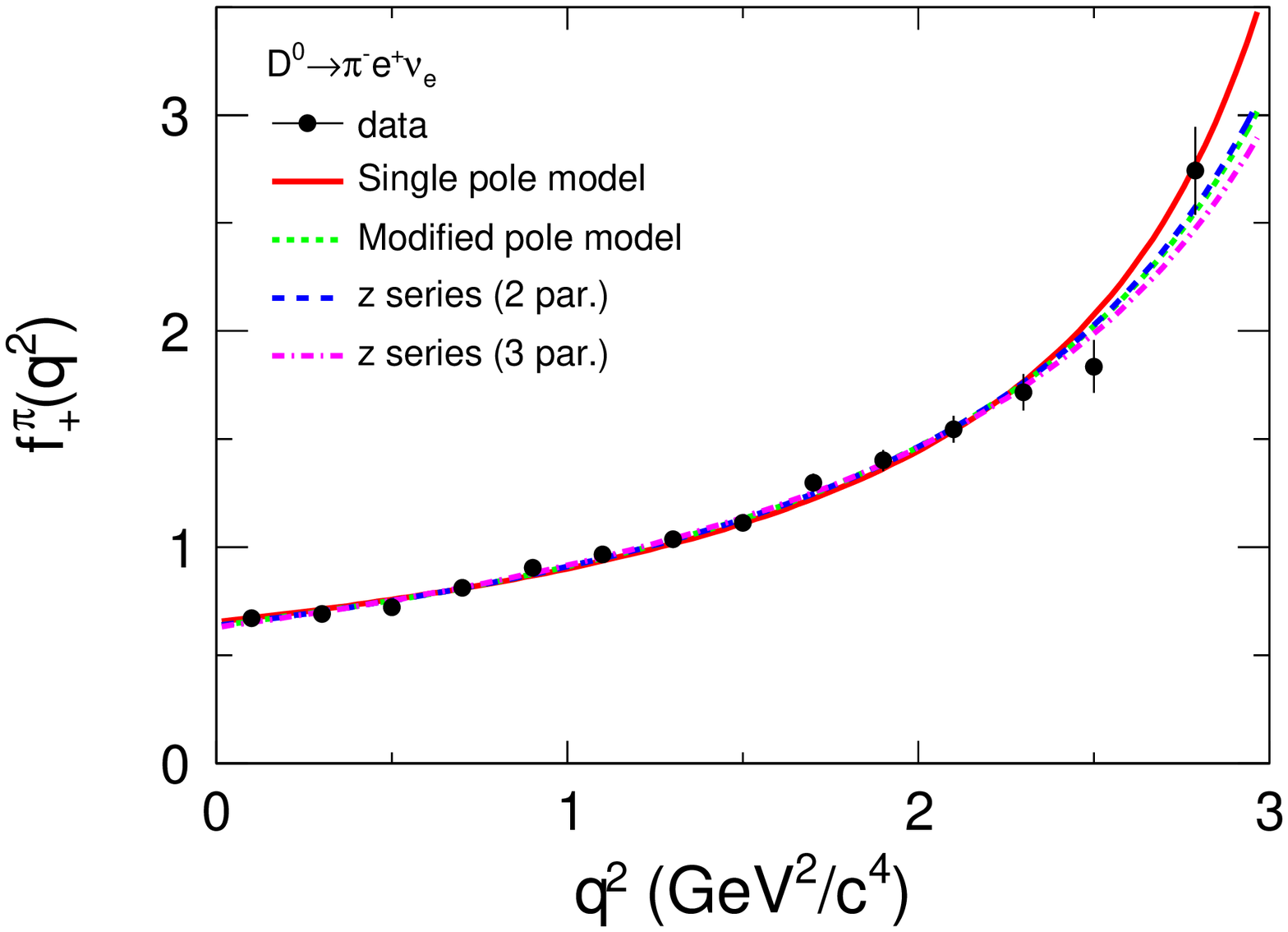}
\caption{Projections on $f_+^{\pi}(q^2)$  for $D^{0}\to \pi^{-}e^{+}\nu_{e}$.}
\label{projections_of_Fits_onto_ff_q2_stat_piev}
\end{minipage}
\end{figure*}

After considering the effects of the systematic uncertainties on the
fitted parameters, we finally obtain the
results of these fits to the partial decay rates with
each form-factor model. The results of these fits are summarized
in Table~\ref{fitresults_on_product_f_and_v_stat_sys_err},
where the first errors are statistical and the second systematic.
The fits to the differential decay rates
for $D^{0}\to K^{-}e^{+}\nu_{e}$ and $D^{0}\to \pi^{-}e^{+}\nu_{e}$
are shown in Figs.~\ref{Form_Factor_Fits_Plots_for_Kev} and
\ref{Form_Factor_Fits_Plots_for_piev}, respectively.

Figures~\ref{projections_of_Fits_onto_ff_q2_stat_Kev} and \ref{projections_of_Fits_onto_ff_q2_stat_piev}
show the projections of
fits onto $f_+(q^2)$ for the $D^{0}\to K^{-}e^{+}\nu_{e}$ and
$D^{0}\to \pi^{-}e^{+}\nu_{e}$ decays, respectively.
In these two figures, the dots with error bars show the measured values of
the form factors, $f_+^{K(\pi)}(q^2)$, which are obtained with
\begin{equation}
f_+^{K(\pi)}(q^2_i)=\sqrt{ \frac{\Delta\Gamma_i}{\Delta q^2_i}
\frac{24\pi^3} {G_F^2 {p^\prime}^3_{K(\pi)} |V_{cs(d)}|^2} }
\end{equation}
in which
\begin{equation}
{p^\prime}^3_{K(\pi)}(i) = \frac{ \int_{q^2_{{\rm min}(i)}}^{q^2_{{\rm max}(i)}} p^3_{K(\pi)}|f_+^{K(\pi)}(q^2)|^2 dq^2}
                           { |f_+^{K(\pi)}(q^2_i) |^2 (q^2_{{\rm max}(i)}-q^2_{{\rm min}(i)}) },
\end{equation}
where
$|V_{cs}|=0.97343\pm 0.00015$ and $|V_{cd}|=0.22522\pm 0.00061$ are taken
from the SM constraint fit~\cite{pdg2014}.
In the calculation of ${p^\prime}^3_{K(\pi)}(i)$, $f_+^{K(\pi)}(q^2)$
and $f_+^{K(\pi)}(q^2_i)$ are computed
using the two parameter series parameterization with the measured parameters.

 \begin{table*}[h!bp]
  \caption{Summary of results of form factor fits to the data.}
  \label{fitresults_on_product_f_and_v_stat_sys_err}
  \begin{ruledtabular}
  \begin{tabular}{lcccc}
  \multicolumn{5}{c}{Single pole model}\\
  Decay mode & $f_+^{K(\pi)}(0)|V_{cs(d)}|$ & $M_{\rm pole}$ (GeV$/c^2$) & & $\chi^2/$d.o.f. \\
  $D^0\to K^- e^+\nu_e$ & $  0.7209\pm0.0022\pm 0.0035$ & $  1.921\pm0.010\pm 0.007$ & & $18.8/16$ \\
  $D^0\to \pi^- e^+\nu_e$ & $  0.1475\pm0.0014\pm 0.0005$ & $  1.911\pm0.012\pm 0.004$ & & $20.0/12$ \\
  \hline
  \multicolumn{5}{c}{Modified pole model}\\
  Decay mode & $f_+^{K(\pi)}(0)|V_{cs(d)}|$ & $\alpha$ & & $\chi^2/$d.o.f. \\
  $D^0\to K^- e^+\nu_e$ & $  0.7163\pm0.0024\pm 0.0034$ & $  0.309\pm0.020\pm 0.013$ & & $20.2/16$ \\
  $D^0\to \pi^- e^+\nu_e$ & $  0.1437\pm0.0017\pm 0.0008$ & $  0.279\pm0.035\pm 0.011$ & & $12.6/12$ \\
  \hline
  \multicolumn{5}{c}{Two-parameter series expansion}\\
  Decay mode & $f_+^{K(\pi)}(0)|V_{cs(d)}|$ & $r_1$ & & $\chi^2/$d.o.f. \\
  $D^0\to K^- e^+\nu_e$ & $  0.7172\pm0.0025\pm 0.0035$ & $ -2.2286\pm0.0864\pm 0.0573$ & & $19.6/16$ \\
  $D^0\to \pi^- e^+\nu_e$ & $  0.1435\pm0.0018\pm 0.0009$ & $ -2.0365\pm0.0807\pm 0.0257$ & & $12.8/12$ \\
  \hline
  \multicolumn{5}{c}{Three-parameter series expansion}\\
  Decay mode & $f_+^{K(\pi)}(0)|V_{cs(d)}|$ & $r_1$ & $r_2$ & $\chi^2/$d.o.f. \\
  $D^0\to K^- e^+\nu_e$ & $  0.7195\pm0.0035\pm 0.0041$ & $ -2.3338\pm0.1587\pm 0.0804$ & $  3.4188\pm3.9090\pm 2.4098$ & $19.1/15$ \\
  $D^0\to \pi^- e^+\nu_e$ & $  0.1420\pm0.0024\pm 0.0010$  & $ -1.8432\pm0.2212\pm 0.0690$ & $ -1.3874\pm1.4615\pm 0.4680$ & $11.9/11$ \\
  \end{tabular}
  \end{ruledtabular}
 \end{table*}

\subsection{Comparison of form-factor parameters in different parameterizations}

For the single pole model, the fits give
\begin{equation}
  M_{\rm pole}^{D \rightarrow K}   = (1.921 \pm  0.010 \pm 0.007)~{\rm GeV}/c^2,
\end{equation}
and
\begin{equation}
  M_{\rm pole}^{D \rightarrow \pi} = (1.911 \pm  0.012 \pm 0.004)~{\rm GeV}/c^2
\end{equation}
for $D^0 \rightarrow K^-e^+\nu_e$ and
$D^0 \rightarrow \pi^-e^+\nu_e$ decays, respectively.
The agreement between the extracted values of pole mass and the expected values
($M_{D^{*+}_{(s)}}$) is extremely poor.
For comparison, Table~\ref{pole_masss_for_simple_pole_model_Kev}
lists the values of the pole mass
$M_{\rm pole}^{D \rightarrow K}$ and $M_{\rm pole}^{D \rightarrow \pi}$
measured in this analysis and those previously measured at other experiments.

\begin{table*}[!hbp]
  \caption{Comparison of measurements of the pole masses $M^{D \rightarrow K}_{\rm pole}$ and $M^{D \rightarrow \pi}_{\rm pole}$.}
  \label{pole_masss_for_simple_pole_model_Kev}
  \begin{ruledtabular}
  \begin{tabular}{lcc}
   Experiment      & $M^{D \rightarrow K}_{\rm pole}$ (GeV$/c^2$)  & $M^{D \rightarrow \pi}_{\rm pole}$ (GeV$/c^2$) \\
   \hline
   Mark III~\cite{mark-iii}                        & $1.80^{+0.50}_{-0.20}\pm0.25$ \\
   E691~\cite{e691}                                & $2.10^{+0.40}_{-0.20}\pm0.20$ \\
   CLEO~\cite{cleo_prd_44_3394}                    & $2.10^{+0.40}_{-0.20}\pm0.25$ \\
   CLEOII~\cite{cleo_ii_plb_317_647}               & $2.00\pm0.12\pm0.18$ \\
   E687 (Tag)~\cite{e687}                          & $1.97^{+0.43}_{-0.22}\pm0.07$ \\
   E687 (Incl)~\cite{e687}                         & $1.87^{+0.11}_{-0.08}\pm0.07$ \\
   CLEO-II~\cite{cleo_prl_94_011802}               & $1.89\pm0.05^{+0.04}_{-0.03}$ & $1.86^{+0.10}_{-0.06}\pm0.05$\\
   FOCUS~\cite{focus_plb_607_233}                  & $1.93\pm0.05\pm0.03$          & $1.91^{+0.15}_{-0.30}\pm0.07$\\
   Belle~\cite{bell_Phys_RevLett_97_p061804_y2006}        & $1.82\pm0.04\pm0.03$          & $1.97\pm0.08\pm0.04$ \\
   \babar~\cite{BaBar_Phys_Rev_D76_p052005_y2007,BaBar_D0topienu}       & $1.884\pm0.012\pm0.015$  & $1.906\pm0.029\pm0.023$\\
   CLEO-c~\cite{cleo-c_Phys_Rev_D79_052010_y2009}  & $1.97\pm0.03\pm0.01$          & $1.95\pm0.04\pm0.02$\\
   CLEO-c~\cite{cleo-c_Phys_Rev_D77_112005_y2008}  & $1.97\pm0.03\pm0.01$          & $1.87\pm0.03\pm0.01$\\
   BESIII (This work)    & $1.921\pm0.010\pm0.007$             & $1.911\pm0.012\pm0.004$\\
  \end{tabular}
  \end{ruledtabular}
 \end{table*}

With the modified pole model, the fits give
\begin{equation}
  \alpha^{D \rightarrow K}   = 0.310 \pm  0.020 \pm 0.013,
\end{equation}
and
\begin{equation}
  \alpha^{D \rightarrow \pi} = 0.279 \pm 0.035 \pm 0.011
\end{equation}
for $D^0 \rightarrow K^-e^+\nu_e$ and
$D^0 \rightarrow \pi^-e^+\nu_e$ decays, respectively.
In the modified pole model (BK parameterization) for the form factors,
$\alpha^{D \rightarrow K}_{\rm BK}$ is expected to be $\sim 1.75$ and
$\alpha^{D \rightarrow \pi}_{\rm BK}$ is expected to be
$\sim 1.34$~\cite{cleo-c_Phys_Rev_D79_052010_y2009}.
Our measured values of $\alpha^{D\rightarrow K}$ and
$\alpha^{D\rightarrow \pi}$ significantly deviate from the values
required by the modified pole model.
Table~\ref{modified_model_form_factor} presents a comparison of our measurements of
these two parameters with those previously measured at other experiments
and the expected values from the Lattice QCD calculations.

\begin{table*}[h!bp]
  \caption{Comparison of measurements of the shape parameters
  $\alpha^{D \rightarrow K}$ and $\alpha^{D \rightarrow \pi}$ in the modified pole model.
  }
\label{modified_model_form_factor}
  \begin{ruledtabular}
  \begin{tabular}{lcc}
   Theory/Experiment      & $\alpha^{D \rightarrow K}$ & $\alpha^{D \rightarrow \pi}$ \\
   \hline
   LQCD~\cite{lqcd_prl_94_011601}         & $0.50\pm0.04\pm0.07$ & $0.44\pm0.04\pm0.07$ \\
   LCSR~\cite{lcsr_nucl40_527}		& $0.07^{+0.15}_{-0.07}$ & $0.11^{+0.11}_{-0.07}$ \\
   FOCUS~\cite{focus_plb_607_233}        & $0.28\pm0.08\pm0.07$ &  \\
   Belle~\cite{bell_Phys_RevLett_97_p061804_y2006}        &    $0.52 \pm 0.08 \pm 0.06$    &  $0.10 \pm 0.21 \pm 0.10$  \\
   CLEO-c($281~\rm pb^{-1}$)(tagged)~\cite{cleo-c_Phys_Rev_D79_052010_y2009}  & $0.21\pm0.05\pm0.02$ & $0.16\pm0.10\pm0.05$ \\
   CLEO-c($281~\rm pb^{-1}$)(untagged)~\cite{cleo-c_Phys_Rev_D77_112005_y2008}& $0.21\pm0.05\pm0.03$ & $0.37\pm0.08\pm0.03$ \\
   CLEO-c($818~\rm pb^{-1}$)~\cite{cleo-c_Phys_Rev_D80_032005_y2009} & $0.30\pm0.03\pm0.01$ & $0.21\pm0.07\pm0.02$ \\
   \babar~\cite{BaBar_Phys_Rev_D76_p052005_y2007,BaBar_D0topienu}        &    $0.377 \pm 0.023 \pm 0.029$    &  $0.268\pm0.074\pm0.059$  \\
   BESIII (This work)    & $0.309 \pm  0.020 \pm 0.013$ & $0.279 \pm 0.035 \pm 0.011$ \\
  \end{tabular}
  \end{ruledtabular}
\end{table*}

\subsection{\boldmath Comparison of the measured $f_+^{K(\pi)}(q^2)$ with LQCD predictions}

\begin{figure}[!hbp]
\centerline{
\includegraphics[width=0.5\textwidth]{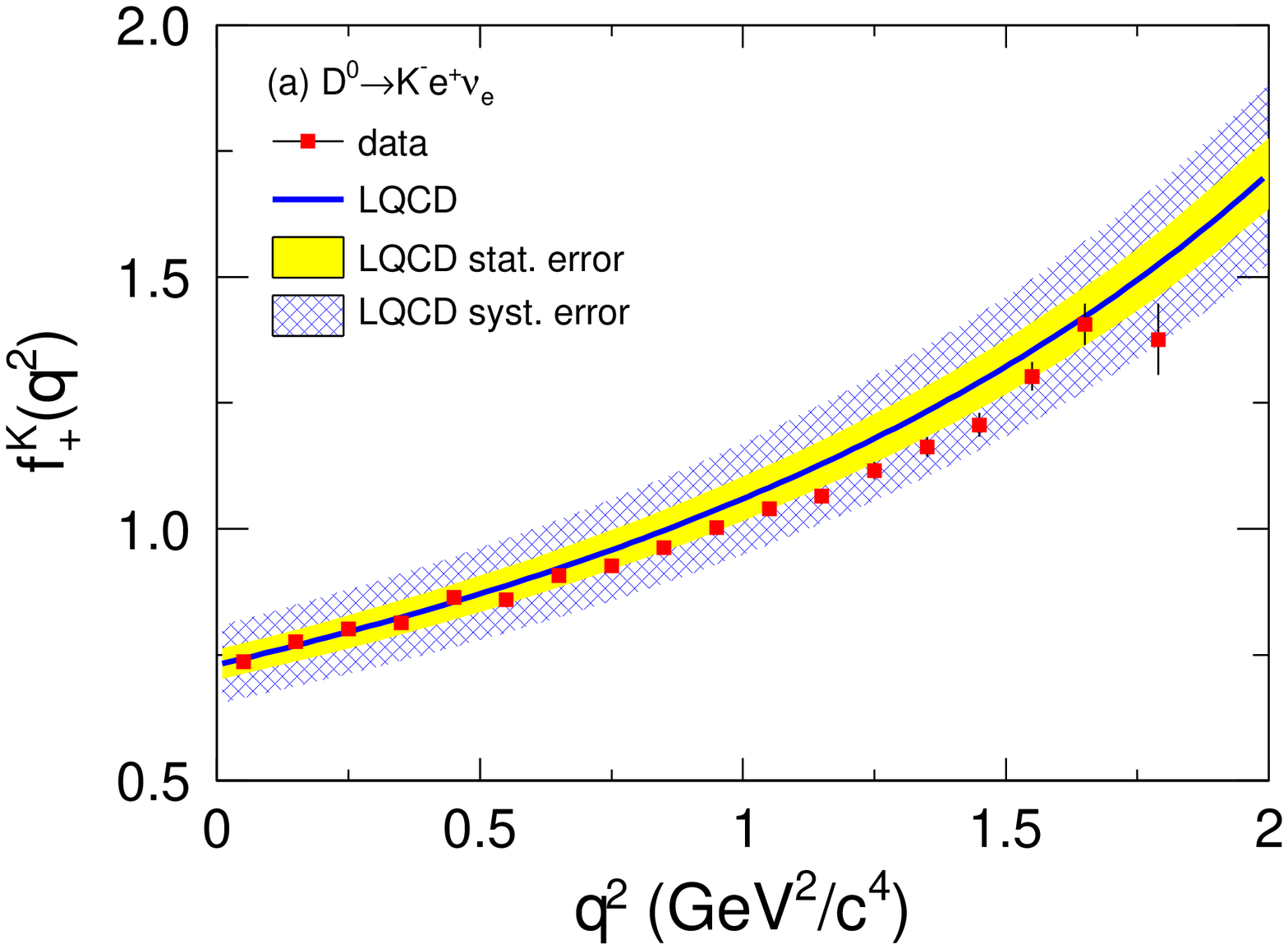}
}
\centerline{
\includegraphics[width=0.5\textwidth]{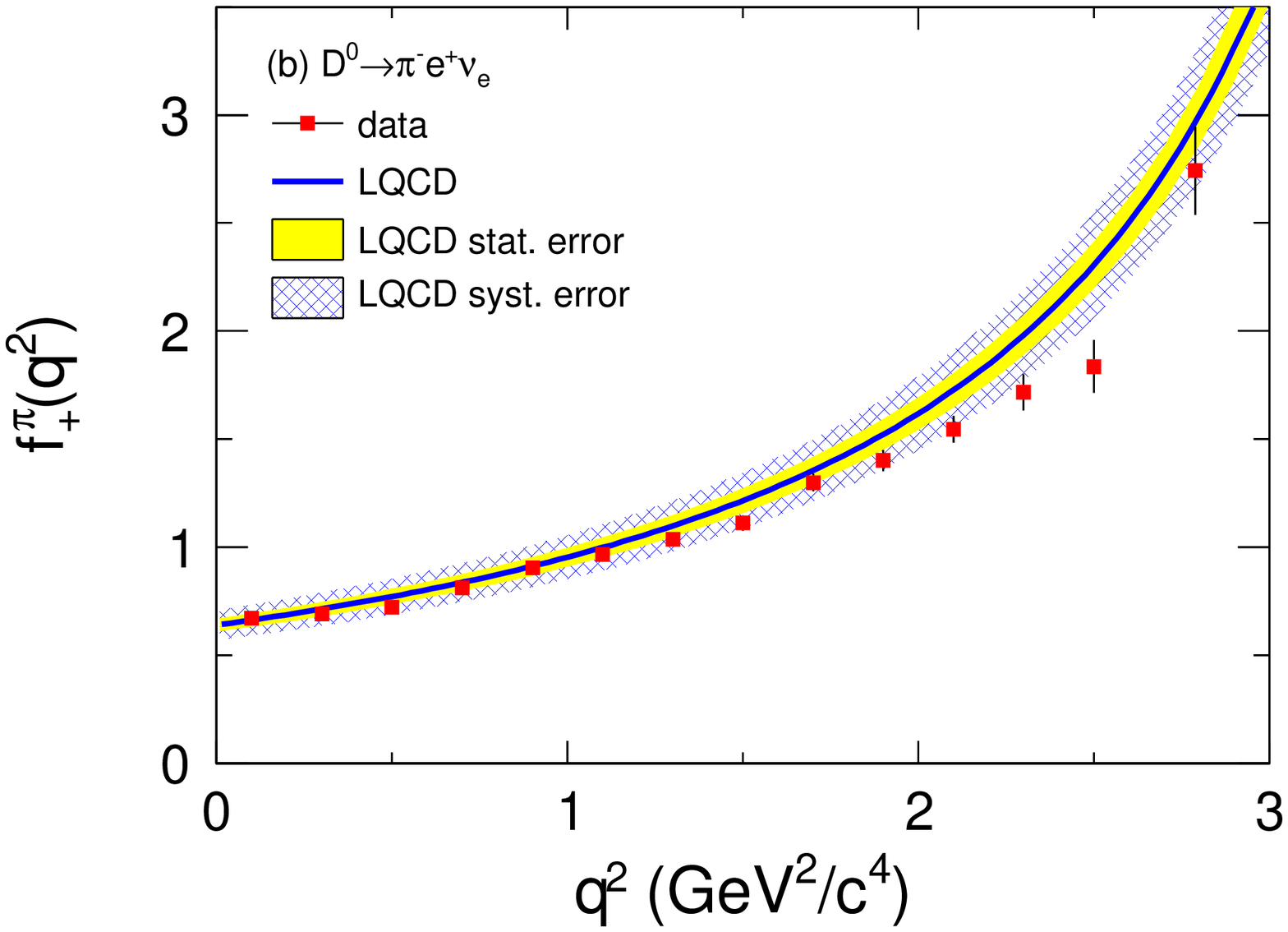}
}
\caption{Comparisons of the measured form factors (squares with error bars)
with the LQCD calculations~\cite{lqcd_prl_94_011601}
(solid
lines present the central values, bands present
the LQCD uncertainties). }
\label{fig_cmp_ff_lqcd}
\end{figure}

Figures~\ref{fig_cmp_ff_lqcd} (a) and (b) show comparisons
between our measured form factors and those calculated in LQCD~\cite{lqcd_prl_94_011601} for
$D^0 \to K^-e^+\nu_e$ and $D^0 \to \pi^-e^+\nu_e$
semileptonic decays, respectively.
From these two figures we find that,
although our measured values of the form factors
$f^K_+(q^2)$ and $f^{\pi}_+(q^2)$
are consistent within uncertainties with the LQCD predictions,
our measured values of the form factors significantly
deviate from
the most probable values calculated in LQCD
in the regions above 0.75 GeV$^2/c^4$ and 1.5 GeV$^2/c^4$ for $D^0 \to K^-e^+\nu_e$
and $D^0 \to \pi^-e^+\nu_e$ decays, respectively.
The precision of the measured $f_+^{K}(q^2)$ and $f_+^{\pi}(q^2)$ is
much higher than that of the LQCD calculations.

\subsection{\boldmath Comparison of measurements of $f^{K}_+(0)$ and $f^{\pi}_+(0)$}

Using the measured $f_+^{K(\pi)}(0)|V_{cs(d)}|$ from the two-parameter
series expansion fits, we obtain
\begin{equation}
\frac{f_+^{\pi}(0)|V_{cd}|}{f_+^{K}(0)|V_{cs}|}=0.2001\pm 0.0026\pm 0.0016,
\label{ratio_products}
\end{equation}
where the first error is statistical and second systematic.
With the values of $|V_{cs(d)}|$ from the SM constraint fit~\cite{pdg2014},
we find
\begin{equation}
\frac{f_+^{\pi}(0)}{f_+^{K}(0)}=0.8649\pm 0.0112\pm 0.0073,
\label{ratio_vpfpi_over_fk}
\end{equation}
where the first error is statistical and second systematic.
This measured ratio, ${f_+^{\pi}(0)}/{f_+^{K}(0)}=0.865\pm 0.013$, is in excellent agreement
with the LCSR calculation of ${f_+^{\pi}(0)}/{f_+^{K}(0)}=0.84\pm 0.04$~\cite{LCSR_1},
but the precision is higher than the LCSR calculation by more than a factor of 3.

Using the $f_+^K(\pi)(0)|V_{cs(d)}|$ values from the two-parameter
series expansion fits and
taking the values of $|V_{cs(d)}|$ from the SM constraint fit~\cite{pdg2014} as inputs,
we obtain the form factors
\begin{equation}
f_+^{K}(0)=0.7368\pm0.0026\pm 0.0036
\label{_fk_series_expansion}
\end{equation}
and
\begin{equation}
f_+^{\pi}(0)=0.6372\pm0.0080\pm 0.0044,
\label{_fpi_series_expansion}
\end{equation}
where the first errors are statistical and the second systematic.

Tables~\ref{tab_cmp_fK_exp} and \ref{tab_cmp_fpi_exp} show the comparisons of our measured form factors with
those measured at other experiments,
for which different form-factor parameterizations and values of
$|V_{cs(d)}|$ have been used.
Our measurements of these two form factors are consistent within
errors with other measurements, but with a higher precision.

\begin{table*}[!hbp]
\caption{Comparison of the form factor $f^{K}_+(0)$ measured at different experiments.}
\label{tab_cmp_fK_exp}
 \begin{ruledtabular}
 \begin{tabular}{lcl}
  Experiment  & $f^{K}_+(0)$ & Form-factor parameterization \\
  \hline
  BES-II~\cite{bes2_plb597_p39_2004} & $0.78\pm0.04\pm0.03 $  &  Single pole model \\
  Belle~\cite{bell_Phys_RevLett_97_p061804_y2006} & $0.695 \pm 0.007 \pm 0.022$ & Modified pole model \\
  \babar~\cite{BaBar_Phys_Rev_D76_p052005_y2007} & $0.727 \pm 0.007 \pm 0.005 \pm 0.007$ & Single pole model and modified pole model \\
  CLEO-c~\cite{cleo-c_Phys_Rev_D80_032005_y2009}  & $0.739 \pm 0.007 \pm 0.005 \pm 0.000$ & Three-parameter series expansion \\
  BESIII (This work)  & $0.7368\pm0.0026\pm 0.0036$ & Two-parameter series expansion \\
 \end{tabular}
 \end{ruledtabular}
\end{table*}

\begin{table*}[!hbp]
\caption{Comparison of the form factor $f^{\pi}_+(0)$ measured at different experiments.}
\label{tab_cmp_fpi_exp}
 \begin{ruledtabular}
 \begin{tabular}{lcl}
  Experiment  & $f^{\pi}_+(0)$ & Form-factor parameterization \\
  \hline
  BES-II~\cite{bes2_plb597_p39_2004}  & $0.73\pm0.14\pm0.06$  &  Single pole model \\
  Belle~\cite{bell_Phys_RevLett_97_p061804_y2006}  & $0.624 \pm 0.020 \pm 0.003$ & Modified pole model \\
  CLEO-c~\cite{cleo-c_Phys_Rev_D80_032005_y2009}  & $0.666 \pm 0.019 \pm 0.004 \pm 0.003$ & Three-parameter series expansion\\
  \babar~\cite{BaBar_D0topienu}       & $0.610\pm0.020\pm0.005$ & Three-parameter series expansion \\
  BESIII (This work) & $0.6372\pm0.0080\pm 0.0044$ & Two-parameter series expansion \\
 \end{tabular}
 \end{ruledtabular}
\end{table*}

\section{\boldmath Extraction of $|V_{cs}|$ and $|V_{cd}|$}
\label{sec:vcq}

\subsection{\boldmath Determination of $|V_{cs}|$ and $|V_{cd}|$}

Using the values for $f_+^K(\pi)(0)|V_{cs(d)}|$ from the two-parameter
$z$-series expansion fits
and in conjunction with
$f_+^{K}(0)=0.747 \pm 0.011 \pm 0.015$~\cite{LQCD_fK}
and $f_+^{\pi}(0)=0.666 \pm 0.020 \pm 0.021$~\cite{LQCD_fpi} calculated in LQCD,
we obtain
\begin{equation}
    |V_{cs}|=0.9601 \pm 0.0033 \pm 0.0047 \pm 0.0239
\end{equation}
and
\begin{equation}
    |V_{cd}|=0.2155 \pm 0.0027 \pm 0.0014 \pm 0.0094,
\end{equation}
where the first uncertainties are statistical,
the second ones systematic,
and the third ones are due to the theoretical uncertainties in
the form factor calculations.

From the measured ratio of $\frac{f_+^{\pi}(0)|V_{cd}|}{f_+^{K}(0)|V_{cs}|}$ given in Eq.(\ref{ratio_products})
together with the LCSR calculation of $f_+^{\pi}(0)/f_+^{K}(0)=0.84\pm 0.04$~\cite{LCSR_1},
we determine
\begin{equation}
\frac{|V_{cd}|}{|V_{cs}|}=0.238\pm 0.004\pm 0.002\pm 0.011,
\end{equation}
where the first error is statistical, the second one systematic,
and the third one is from LCSR normalization.

\subsection{\boldmath Comparison of $|V_{cs}|$ and $|V_{cd}|$}

Table~\ref{cmp_Vcs} and Table~\ref{cmp_Vcd} give comparisons of our measured $|V_{cs}|$ and $|V_{cd}|$ with those
measured at other experiments.
Our measurements of $|V_{cs}|$ and $|V_{cd}|$
are of higher precision than previous results from both $D$ meson decays
and $W$ boson decays.

\begin{table*}
\caption{Comparison of $|V_{cs}|$ measurements.}
\label{cmp_Vcs}
\begin{ruledtabular}
  \begin{tabular}{lcl}
   Experiment & $|V_{cs}|$ &    Note \\
   \hline
   PDG2014~\cite{pdg2014} &  $0.986\pm 0.016$                   & Using $D_s^+\rightarrow \ell^+\nu_\ell$, $D^0 \rightarrow K^-\ell^+\nu_\ell$ and $D^+\to\bar K^0e^+\nu_e$  \\
   PDG2014~\cite{pdg2014} &  $1.008\pm 0.021$                   & Using $D^+_s \rightarrow \ell^+\nu_\ell$  \\
   PDG2014~\cite{pdg2014} &  $0.953\pm 0.008\pm0.024$                      & Using $D^0 \rightarrow K^-\ell^+\nu_\ell$ and $D^+\to\bar K^0e^+\nu_e$ \\
   PDG2014~\cite{pdg2014} &  $0.97343\pm 0.00015$                & Global fit in the Standard Model \\
   PDG2006~\cite{pdg2006}  &  $0.94^{+0.32}_{-0.26}\pm 0.14$      & Using $W$ boson decay \\
   BES-II~\cite{bes2_plb597_p39_2004} &  $1.00\pm 0.05\pm 0.11$              & Using $D^0 \rightarrow K^-e^+\nu_e$ \\
   CLEO-c~\cite{cleo-c_Phys_Rev_D80_032005_y2009} &  $0.985\pm 0.009\pm 0.006\pm0.103$   & Using $D^0 \rightarrow K^-e^+\nu_e$ and $D^+\to\bar K^0e^+\nu_e$\\
   BESIII (This work)          &  $0.9601 \pm 0.0033 \pm 0.0047 \pm 0.0239$  & Using $D^0 \rightarrow K^-e^+\nu_e$   \\
  \end{tabular}
\end{ruledtabular}
\end{table*}

\begin{table*}
\caption{Comparison of  $|V_{cd}|$ measurements.}
\label{cmp_Vcd}
\begin{ruledtabular}
  \begin{tabular}{lcl}
   Experiment & $|V_{cd}|$ &  Note \\
   \hline
   PDG2014~\cite{pdg2014}  &  $0.225\pm 0.008$  & Using $D^0\to\pi^-\ell^+\nu_\ell$,
   $D^+\to\pi^0e^+\nu_e$ and neutrino interactions  \\
   PDG2014~\cite{pdg2014}  &  $0.220\pm 0.006\pm0.010$  & Using $D^0\to\pi^-\ell^+\nu_\ell$ and $D^+\to\pi^0e^+\nu_e$ \\
   PDG2014~\cite{pdg2014}  &  $0.230\pm 0.011$  & Using neutrino interactions  \\
   PDG2014~\cite{pdg2014} &  $0.22522\pm 0.00061$                & Global fit in the Standard Model \\
   CLEO-c~\cite{cleo-c_Phys_Rev_D80_032005_y2009}        &  $0.234\pm 0.007\pm 0.002 \pm 0.025$  &   Using $D^{0} \rightarrow \pi^{-} e^+\nu_e$ and $D^+\to\pi^0e^+\nu_e$ \\
   BESIII~\cite{RongG_Charm2012}                        & $0.2210\pm 0.0058\pm 0.0047$    & Using $D^+ \rightarrow \mu^+\nu_\mu$  \\
   \babar~\cite{BaBar_D0topienu}  & $0.206\pm0.007\pm0.009$ & Using $D^0 \rightarrow \pi^-e^+\nu_e$  \\
   BESIII  (This work)               &  $0.2155 \pm 0.0027 \pm 0.0014 \pm 0.0094$   & Using $D^0 \rightarrow \pi^-e^+\nu_e$  \\
  \end{tabular}
\end{ruledtabular}
\end{table*}

Table~\ref{cmp_Vcd_over_Vcs} gives a comparison of our measured ${|V_{cd}|}/{|V_{cs}|}$ with
the one measured by CLEO-c~\cite{cleo-c_Phys_Rev_D80_032005_y2009}
and the world average calculated with $|V_{cd}|$ and $|V_{cs}|$ given
in PDG2014~\cite{pdg2014}.
Our measurement of the ratio is in excellent agreement with the world average.

\begin{table*}
\caption{Comparison of  $|V_{cd}|/|V_{cs}|$ measurements.}
\label{cmp_Vcd_over_Vcs}
\begin{ruledtabular}
  \begin{tabular}{lcl}
   Experiment & $|V_{cd}|/|V_{cs}|$ &  Note \\
   \hline
   PDG2014~\cite{pdg2014}  &  $0.228\pm 0.009$  & Using $|V_{cd}|=0.225\pm 0.008$ and $|V_{cs}|=0.986\pm 0.016$   \\
   CLEO-c~\cite{cleo-c_Phys_Rev_D80_032005_y2009}        &  $0.242\pm 0.011\pm 0.004 \pm 0.012$
                                                         &   Using $D \rightarrow \pi e^+\nu_e$ and $D \rightarrow K e^+\nu_e$  \\
   BESIII (This work)     &  $0.238\pm 0.004\pm 0.002\pm 0.011$     & Using $D^0 \rightarrow \pi^-e^+\nu_e$ and $D^0 \rightarrow K^-e^+\nu_e$ \\
  \end{tabular}
\end{ruledtabular}
\end{table*}

\section{Summary}
\label{sec:sum}

In summary, by analyzing about 2.92~fb$^{-1}$ data collected at 3.773~GeV
with the BESIII detector operated at the BEPCII collider,
the semileptonic decays of $D^0 \rightarrow K^-e^+\nu_e$
and  $D^0 \rightarrow \pi^-e^+\nu_e$
have been studied. From a total of $2793317 \pm 3684$
single $\bar D^0$ tags,
$70727.0 \pm 278.3$ $D^0 \to K^-e^+\nu _e$ and
$6297.1 \pm 86.8$  $D^0 \to \pi ^-e^+\nu_e$ signal events
are observed in the system recoiling against the single $\bar D^0$
tags. These yield the decay branching fractions
$$\mathcal B(D^0 \to K^-e^+\nu_e)=(3.505\pm 0.014 \pm 0.033)\%$$
and
$$\mathcal B(D^0 \to \pi^-e^+\nu_e)=(0.295\pm 0.004\pm 0.003)\%.$$

Using these samples of $D^0 \rightarrow K^-e^+\nu_e$
and $D^0 \rightarrow \pi^-e^+\nu_e$ decays,
we study the form factors as a function of the squared four-momentum
transfer $q^2$ for these two decays.
By fitting the partial decays rates, we obtain the parameter values
for several different form-factor functions.
For the physical interpretation of the shape parameters in the single pole
and modified pole models, the values of the parameters obtained from
our fits significantly deviate from those expected by these models.
This means that the data do not support the physical interpretation of the shape parameter in those models.
We choose the values of
$f_+^K(0)|V_{cs}|$ and $f_+^{\pi}(0)|V_{cd}|$
obtained with the two-parameter series expansion as our main result.
In this case, we obtain the form factors
$$f^K_+(0) = 0.7368\pm0.0026\pm 0.0036$$
and
$$f^{\pi}_+(0) = 0.6372\pm0.0080\pm 0.0044.$$
Furthermore, using the form factors calculated in recent LQCD
calculations~\cite{LQCD_fK,LQCD_fpi}, we obtain the CKM matrix elements
$$|V_{cs}|=0.9601 \pm 0.0033 \pm 0.0047 \pm 0.0239$$
and
$$|V_{cd}|=0.2155 \pm 0.0027 \pm 0.0014 \pm 0.0094,$$
where the errors are dominated by the theoretical
uncertainties in the form factor calculations.
Our measurement of the product $f_{+}^K(0)|V_{cs}|=0.7172\pm0.0025\pm 0.0035$
($f_{+}^{\pi}(0)|V_{cd}|=0.1435\pm0.0018\pm 0.0009$) 
is the most precise to date and 
would give more precise value
of $|V_{cs}|$ ($|V_{cd}|$) with its precision increasing to $0.6\%$ ($1.4\%$)
when the uncertainty of the value of the related form factor calculated in LQCD can be ignored.

Our measurements of the branching fractions, the form-factor parameters
and the shapes of the form factor $f_+^{K(\pi)}(q^2)$ as a function of $q^2$
for $D^0 \to K^-e^+\nu_e$ and $D^0 \to \pi^- e^+\nu_e$ decays
are all the most precise to date. These precise measurements
of $f_+^{K}(q^2)$, $f_+^{\pi}(q^2)$, $f_+^K(0)$, $f_+^{\pi}(0)$
and $f_+^{\pi}(0)/f_+^K(0)$ are in good agreement
with the LQCD calculations of the form factors and the LCSR calculations
of the ratio of the form factors, but have higher precision than those calculated in theories
based on QCD, and therefore will allow incisive tests of any future theoretical calculations.

\begin{acknowledgments}
The BESIII collaboration thanks the staff of BEPCII and the IHEP computing center for their strong support.
This work is supported in part by National Key Basic Research Program of China under Contracts
No. 2009CB825204, No. 2015CB856700; National Natural Science Foundation of China (NSFC) under Contracts
Nos. 10935007, 11125525, 11235011, 11322544, 11335008, 11425524;
the Chinese Academy of Sciences (CAS) Large-Scale Scientific Facility Program;
Joint Large-Scale Scientific Facility Funds of the NSFC and CAS under Contracts Nos. 11179007, U1232201, U1332201;
CAS under Contracts Nos. KJCX2-YW-N29, KJCX2-YW-N45; 100 Talents Program of CAS;
INPAC and Shanghai Key Laboratory for Particle Physics and Cosmology;
German Research Foundation DFG under Contract No. Collaborative Research Center CRC-1044;
Istituto Nazionale di Fisica Nucleare, Italy; Ministry of Development of Turkey under Contract No. DPT2006K-120470;
Russian Foundation for Basic Research under Contract No. 14-07-91152;
U. S. Department of Energy under Contracts Nos. DE-FG02-04ER41291, DE-FG02-05ER41374, DE-FG02-94ER40823, DESC0010118;
U.S. National Science Foundation; University of Groningen (RuG) and the Helmholtzzentrum fuer Schwerionenforschung GmbH (GSI), Darmstadt;
Institute for Basic Science, Korea, Project code IBS-R016-D1; the Swedish Research Council.
\end{acknowledgments}

\end{document}